\documentclass[trackchanges,twocolumn]{aastex7}
\usepackage{amsmath}
\usepackage{changepage,graphicx}
\usepackage{amsfonts}

\begin{document}

\title{Simulations of structured upflows from plumes and their connection to the solar wind}

\author[orcid=0000-0002-3402-273X]{K. Bora}
\affiliation{Max-Planck Institute for Solar System Research, 37077 G{\"o}ttingen, Germany}
\email[show]{bora@mps.mpg.de}  

\author[orcid=0000-0002-9270-6785]{L. P. Chitta} 
\affiliation{Max-Planck Institute for Solar System Research, 37077 G{\"o}ttingen, Germany}
\email{chitta@mps.mpg.de} 

\author[orcid=0000-0001-5494-4339]{Y. Chen}
\affiliation{Max-Planck Institute for Solar System Research, 37077 G{\"o}ttingen, Germany}
\email{cheny@mps.mpg.de} 

\author[orcid=0000-0003-1670-5913]{D. Przybylski}
\affiliation{Max-Planck Institute for Solar System Research, 37077 G{\"o}ttingen, Germany}
\email{przybylski@mps.mpg.de} 

\author[orcid=0000-0002-1089-9270]{D.I.Pontin}
\affiliation{School of Information and Physical Sciences, University of Newcastle, University Drive, Callaghan, NSW 2308, Australia}
\email{David.Pontin@newcastle.edu.au} 

\author[orcid=0009-0007-5692-4922]{N. Panyam}
\affiliation{Institut für Astrophysik und Geophysik, Georg-August-Universit{\"a}t, 37077 G{\"o}ttingen, Germany}
\email{nikil.panyam@stud.uni-goettingen.de}

\begin{abstract}
Small-scale transient jetlet activity and associated upflows from coronal hole plumes are potential sources of the solar wind. To elucidate the magnetic origins and driving mechanisms of such upflows, we perform three-dimensional radiative magnetohydrodynamic simulations using the MURaM code, spanning from the upper convection zone to the low corona. We synthesize Fe\,{\sc x} 174 {\AA} emission to capture the plume evolution comparable to observations, examining underlying plasma flows, thermal structures, and magnetic topologies. We identify a pronounced transition from cool downflows in the lower atmosphere to hot upflows in the corona at interface between plume-rooted like-polarity flux concentrations. These upflows are threaded by a complex, filamentary network of Quasi-Separatrix Layers (QSLs)--- a topology distinct from standard interchange reconnection scenarios. The domain-averaged mass flux over a 38-minute interval ranges from $10^{-9}$ to $10^{-8}\,\mathrm{g\,cm^{-2}\,s^{-1}}$, substantially exceeding observed solar-wind loss rates. Our results demonstrate that highly structured plasma outflows are channeled along strong QSLs at open–open field boundaries, providing a pathway to sustain the solar wind from coronal-hole plumes without requiring interchange reconnection triggered by opposite-polarity flux emergence.
\end{abstract}


\section{Introduction}
Coronal holes are magnetically open regions of the solar atmosphere, characterized by their low density, reduced temperatures, and dark appearance in extreme ultraviolet (EUV) and X-ray wavelengths \citep{cranmer2009, wang2009}. These regions are widely recognized as sources of the high-speed solar wind, with plasma escaping along open magnetic field lines into the heliosphere \citep{cranmer2002}. Understanding the physical mechanisms that govern mass and energy outflow from coronal holes is a fundamental goal in heliophysics, with implications for both solar wind generation and space weather forecasting.

Several observed phenomena within coronal holes have been  proposed to contribute to the nascent solar wind which include: (1) small-scale, transient jet-like features known as jetlets emerging at the bases of plumes and interplume regions \citep{raouafi2014, Kumar2022, chitta2023, chitta2025}, (2) persistent, quasi-steady hot upflows observed throughout coronal hole interiors \citep{hassler1999, tu2005, tian2011}, and (3) Alfv\'enic waves \citep{depontieu2007, McIntosh2011}.

Plumes, which appear bright in the EUV, are radially extended structures superimposed on the darker background of coronal holes. They are anchored in network magnetic flux concentrations on the solar surface \citep{gabriel2009}. Plumes exhibit dynamic, fine-scale filamentary substructure, and are frequently populated by collimated propagating intensity disturbances, termed jetlets, at their bases. Observations hint at a complex interplay between magnetic topology, plasma dynamics, and energy release \citep{raouafi2014, uritsky2021}. The association between plume activity and the solar wind has not yet been fully established. The details of processes driving the plume activity also remain less understood, although studies indicate that it is potentially caused by the interchange reconnection between the open and closed field configuration at the plume base \citep{fisk2005, Kumar2022}.

Jetlets, often interpreted as signatures of magnetic reconnection between newly emerging flux and pre-existing open field lines, are thought to episodically inject mass and energy into the corona. Jetlets emerging from plumes have base widths ranging from a few hundred kilometers to a few thousand kilometers \citep[][]{Kumar2022,chitta2023}.

Beyond discrete jet events, sustained upflows in coronal holes have also been reported in spectroscopic observations, particularly in emission lines formed in the transition region and low corona. Data from the SUMER instrument onboard SOHO have revealed a transition from cool downflows to hot upflows with increasing height \citep{dammasch1999}, often concentrated at magnetic network boundaries \citep{hassler1999}. These observations suggest that persistent mass loading may occur in narrow, structured lanes, potentially contributing to the continuous component of the solar wind.

A key theoretical framework invoked to explain these phenomena involves magnetic reconnection, particularly interchange reconnection between open and closed field regions \citep{fisk2005}. In this context, newly emerging opposite-polarity flux reconnects with ambient open field lines, releasing energy and accelerating plasma along open pathways. However, there are also studies that drew attention to magnetic configurations lacking null points, focusing instead on the role of Quasi-Separatrix Layers (QSLs) — regions where magnetic field line connectivity changes rapidly without discontinuity \citep{priest1995, demoulin1996}, to explain coronal phenomena. QSLs have been proposed as sites for reconnection-driven energy release and as potential conduits for guiding plasma into the corona. 

Beyond interchange reconnection at null points, several studies have long emphasized the role of component reconnection between adjacent open flux tubes in coronal holes. In particular, \citet{Priest2002} outlined multiple ways in which the magnetic carpet could energize the corona, including component reconnection driven by braiding and footpoint motions. At the same time, photospheric buffeting is known to launch Alfv\'enic waves into the corona, where nonlinear interactions can lead to turbulence, current sheets, and subsequent reconnection \citep{Matthaeus1999, Leamon2000, Dmitruk2002, Servidio2010, Rappazzo2012}. All these  processes can potentially heat the material in open-field regions and accelerate the fast solar wind. Our work builds on this broader context by focusing on how quasi-separatrix layers (QSLs) formed between like-polarity open flux tubes in plumes can channel structured plasma upflows, providing a more specific topological framework within which such component reconnection may operate.

Numerical modeling, particularly with three-dimensional (3D) radiative magnetohydrodynamic (MHD) codes, provides a critical tool for investigating the magnetic origins of upflows emerging from plumes, in a controlled, high-resolution setting. In this context, the MURaM code \citep{voegler2005} has emerged as a powerful platform for simulating solar atmosphere dynamics, capable of resolving fine magnetic structures and capturing the interplay of convection, radiation, and reconnection.

In this study, we focus on the detailed dynamics of a coronal hole plume simulated using the MURaM code, emphasizing the role of magnetic topology, small-scale structures, and potential plasma outflow channels. By leveraging a spatial resolution that exceeds that of current coronal EUV observations, we aim to explore the substructure of coronal plumes and investigate the conditions under which mass is transported from the lower solar atmosphere into the corona. 


\begin{figure*}[ht!]
   \centering
   \includegraphics[width=\linewidth, trim={0 1.8cm 0 0},clip]{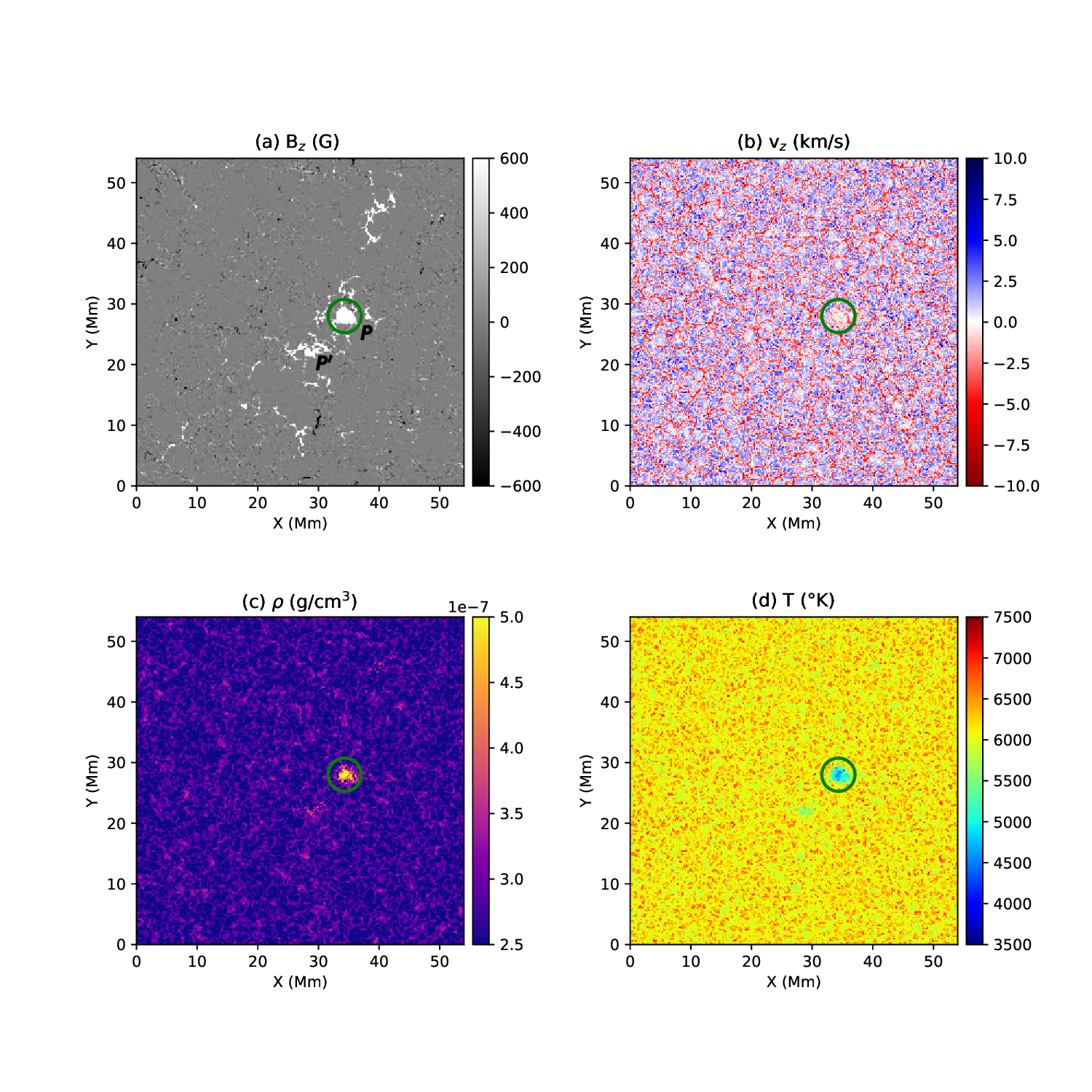}
      \caption{MHD simulations of a plume. A snapshot of physical properties of the plume as defined by the concentrated positive polarity flux tube at the photosphere ($\tau_{500nm}\approx 1$) in our model. The full horizontal extent of the computational domain is shown. Panel (a): vertical component of the magnetic field, representing the photospheric magnetogram where P  and P$'$ mark the plume and neighboring positive polarity patches respectively. Panel (b): vertical component of the flow velocity exhibiting the granulation pattern. Panel (c): density. Panel (d): temperature. The feature marked by green circle in all the panels is the photospheric footpoint of the plume. An animation of this figure is available \url{online}, illustrating the temporal evolution of all four quantities shown in panels (a)–(d) over 57 minutes of solar time at a cadence of 3.8 minutes. The real-time duration of the video is 8~s.
      See Section \ref{subsec3p1} for details.
      }
         \label{fig1}
   \end{figure*}

\section{Numerical Methods}
We perform a three-dimensional radiative magnetohydrodynamic (MHD) simulation using the coronal extension of the MURaM code \citep{vogler2007,rempel2017}, which enables self-consistent modeling of the solar atmosphere from the upper convection zone to the corona. This advanced implementation includes a gray local-thermal-equilibrium (LTE) radiative transfer scheme in the lower atmosphere, optically thin radiative losses in the upper atmosphere, and field-aligned Spitzer thermal conduction \citep{spitzer1962}.

To simulate a self-consistent evolution of a plume within a coronal hole, we employ a two step procedure to construct the initial state of the simulation. In the first step, we create a coronal-hole-like configuration by imposing a uniform vertical magnetic field of 5\,G in a pre-evolved quiet-Sun model, similar to that in \citet{chen2021}. This set up is then evolved self-consistently for several solar hours during which the uniform magnetic field is advected into the intergranular lanes and restructured by the magnetoconvection, yielding mixed-polarity magnetic field, but preserving the net signed flux. 

In the second step, we introduce a vertical magnetic flux tube near the center of the computational domain prepared in the first step, specifically at a site characterized by strong convective downflows. Details of the flux-tube insertion and the subsequent extension of the computational box into the corona are provided in Appendix~\ref{ap1}. Our analysis focuses on plasma dynamics during the final solar hour of the simulation period; snapshots illustrating these dynamics are presented in the following section. 

The computational domain spans $54 \times 54 \times 51$~Mm\textsuperscript{3}, extending from approximately 20~Mm below the photosphere to 30~Mm into the corona. The domain is resolved with $1024 \times 1024 \times 2550$ grid points, corresponding to a spacing of roughly 52~km in the horizontal ($x$, $y$) and 20~km in the vertical ($z$) direction. This setup will capture the granular evolution and the subsequent magnetic field dynamics, at spatial resolutions comparable to observations.

Boundary conditions are periodic in the horizontal directions. The top boundary is semi-transparent---open for outflows and closed for inflows. At the top boundary the magnetic field is specified by a potential boundary condition. This allows plasma and MHD waves to exit the domain without significant reflection, facilitating a natural evolution of coronal structures. In practice, this is implemented by prescribing the zero normal derivative condition on flow velocity. The lower boundary condition is open to inflows and outflows (for more details, see \citet{rempel2014, rempel2017}). 
 
\begin{figure*}[ht!]
   \centering
   \includegraphics[width=\linewidth, trim={0 4cm 0 0},clip]{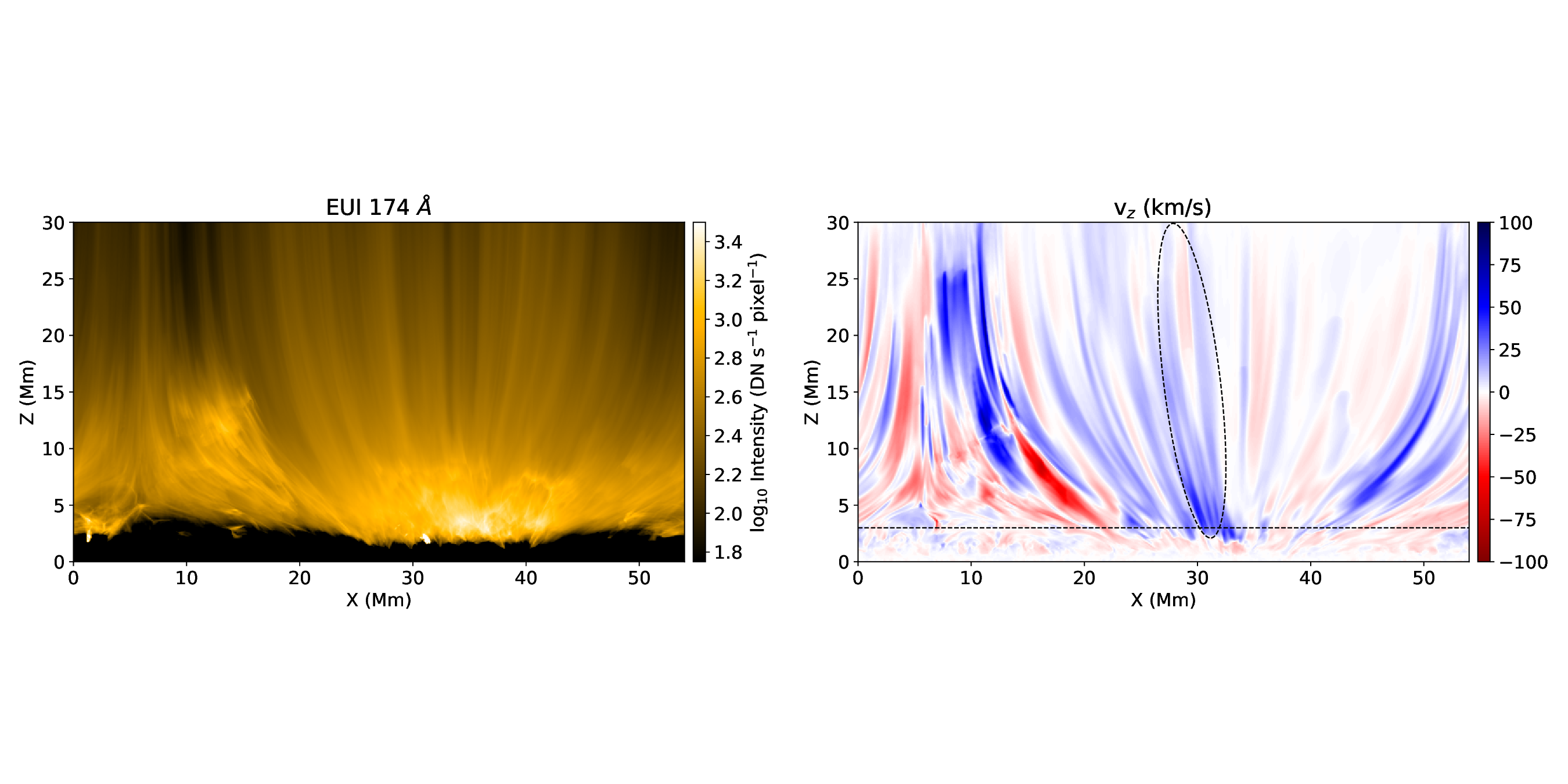}
   \caption{Limb or side-view of the coronal emission and flow structure showing the vertical expansion of the plume in our 3D MHD model. Left: coronal emission synthesized in EUI 174 {\AA} filter, integrated along $y-$direction (displayed on the $xz-$plane) in the model. Right: vertical component of the flow velocity at $y \approx$ 23 Mm, on $xz-$plane. The ellipse on $v_{z}$ slice marks the upflows channeled at the boundary of plume.} 
         \label{fig2}
\end{figure*}

\begin{figure*}[ht!]
   \centering
   \includegraphics[width=\linewidth, trim={0 2.5cm 0 0},clip]{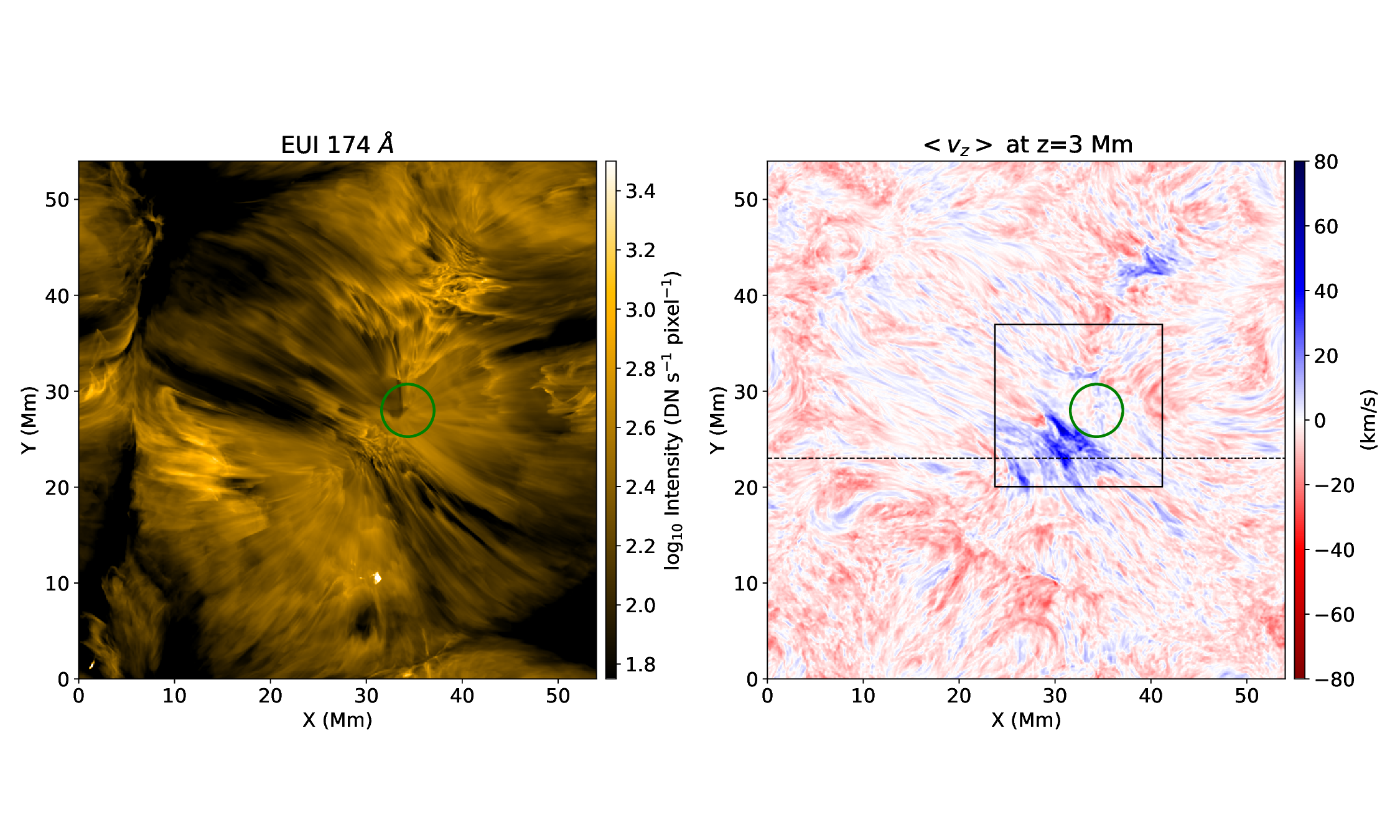}
   \caption{Disk-center or top-down view of the coronal emission and flow structure in our 3D MHD model. Left: coronal emission synthesized in EUI 174 {\AA} filter, integrated along vertical direction ($z$) in the model. Right: vertical component of the flow velocity, time-averaged over approximately 18 minutes at 3 Mm height above the photosphere. The upflow structure at the plume boundary is enclosed within black rectangle. The green circles (same as Figure \ref{fig1}) in both the images mark the location of introduced plume. The dashed line is reference for a slice of $v_{z}$ map in Figure~\ref{fig3}. See Section \ref{subsec3p2} for details. An animation combining the left panels of  the current figure and Figure~\ref{fig2} is available \url{online}. It illustrates the temporal evolution of synthesized coronal emission in EUI 174~{\AA} filter, spanning approximately 38 minutes of solar time with a cadence of 22s. The real-time duration of the animation is 7~s. See Section \ref{subsec3p2} for details.} 
         \label{fig3}
\end{figure*}

\section{Results}

The numerical simulations of coronal hole plumes conducted in this study provide detailed insights into the dynamic processes governing the structure and evolution of plumes in the solar atmosphere. Key findings are summarized below:

\subsection{Physical Properties of the Introduced Pore/Open Flux Tube on the Photosphere}
\label{subsec3p1}
Figure~\ref{fig1} illustrates the spatial distribution of key physical properties of the introduced pore/open flux tube at the corrugated optical depth layer, corresponding to a continuum optical depth at 500\,nm. Each panel provides a snapshot of the full horizontal extent of the computational domain ($\approx$54 Mm$\times$54 Mm), highlighting distinctive physical characteristics associated with the magnetic structure.

The vertical magnetic field component ($B_z$), clearly shows the originally introduced open magnetic flux tube characterized by a strong, concentrated unipolar magnetic region (Figure~\ref{fig1}a).

The granulation pattern can be discerned by the vertical component of the flow velocity ($v_z$), indicative of vigorous convection typical of photospheric plasma (Figure~\ref{fig1}b). In particular, convection is substantially suppressed ($v_{z}\approx0$) at the location of the intense magnetic flux tube, emphasizing the magnetic inhibition of convective motions within strong magnetic fields. 

An enhanced plasma density is evident at the site of the magnetic pore, reflecting the Wilson depression phenomenon (Figure~\ref{fig1}c). Within magnetic regions, the $\tau=1$ surface is physically deeper compared to nonmagnetic regions as a result of magnetic pressure support, positioning it within denser layers of the photosphere. 

The magnetic pore region is cooler relative to the surrounding nonmagnetic plasma (Figure~\ref{fig1}d). The cooler temperature results from reduced convective energy transport as a result of magnetic field suppression of convection, reinforcing the thermal distinction between magnetic pores and the ambient plasma environment. Together, these visualizations underscore how the plume magnetic field can affect local photospheric conditions, manifesting as clear perturbations in magnetic intensity, flow velocities, density stratification, and temperature distributions.  

Notably, panel (a) shows the positive‐polarity concentrations away from originally introduced open flux tube, clustered around the green‐circled region. The parent polarity has been fragmented and the flux concentrations are advected  away by convective flows. Notably, the convection is not suppressed at the polarities away from original one but we note the temperature is low at the left concentration. 

The region, marked by a green circle, is distinct compared to the ambient, weaker magnetic field regions with maximum $B_z\approx1800$~G, size $\approx$ 3~Mm and the flux $\approx$ 4.375$\times$10$^{19}$~Mx at z=0 (photospheric reference slice in the simulation domain). The strong magnetic field within this flux tube significantly influences plasma dynamics and thermodynamics.

\begin{figure*}[ht!]
   \centering
   \includegraphics[width=\linewidth]{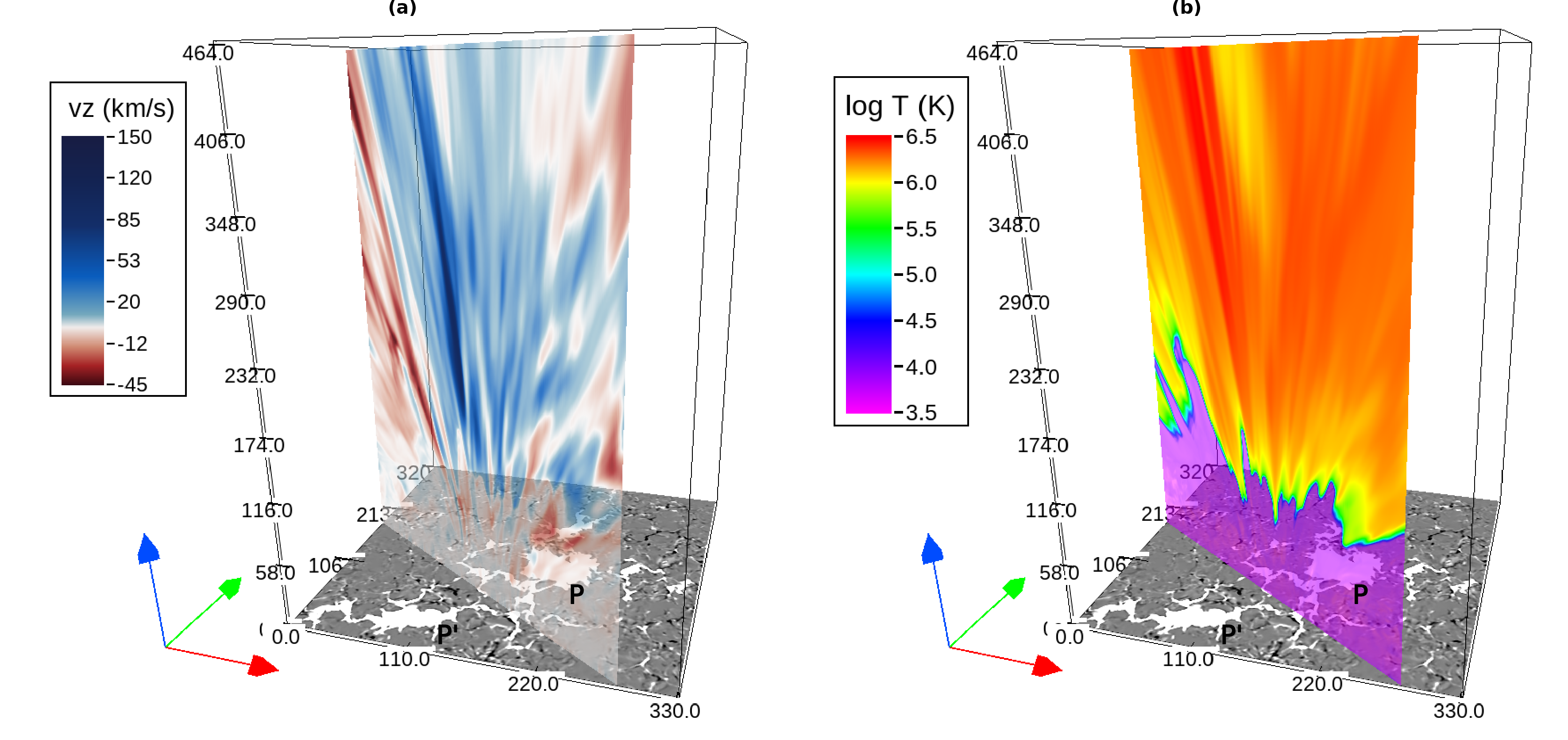}
   \caption{Visualizing thermodynamics. Panel (a): vertical component of the velocity on a plane passing between two positive polarity patches of plume P and neighboring polarity P$'$ (blue denotes upflowing plasma and red denotes downflowing plasma). Panel (b): temperature (in $\log$ scale) on the same vertical plane. A transition from cool downflowing plasma to hot upward moving plasma is visible above a certain height. In each panel, the red, green, and blue arrows on  bottom left indicate the $x$, $y$, and $z$-axes of the 3D Cartesian coordinate system, respectively. The bottom boundary is $B_z$ saturated between $\pm$400 G. An animation of this figure is available \url{online}, showing the temporal evolution of the displayed quantities over 38 minutes of solar time with a cadence of 3.8 minutes. The animation runs for 5~s in real time. See Section \ref{subsec3p2} for details.}
         \label{fig4}
\end{figure*}

\begin{figure*}[ht!]
   \centering
   \includegraphics[width=\linewidth]{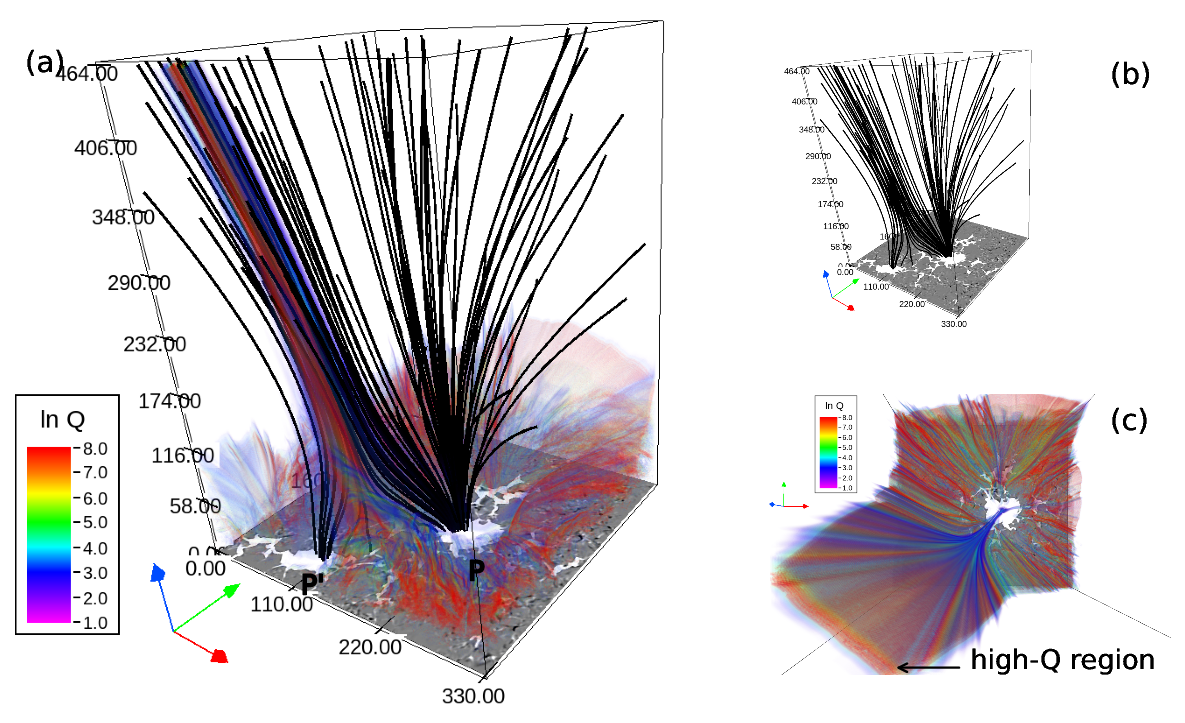}
   \caption{QSLs and magnetic structures shaping the flow and temperature profiles between two positive magnetic polarities P and P$'$ which form the interface between two open field regions. Panel (a): Direct volume rendering of the $\ln Q$ distribution, overlaid with magnetic field lines (black) originating from the two adjacent positive polarities, for the same field of view as in Figure~\ref{fig4}. The surface corresponding to the very high value $\ln Q = 8$ (red) is located at the midplane, marking the interface formed by field lines from the two open field regions. Panel (b): Side view showing the inverted Y-shaped magnetic structure that forms the interface between the two positive polarities/open field regions. Panel (c): Top-down view of the QSLs displayed in panel (a), emphasizing the intricate fine structures. See Section~\ref{subsec3p3} for further details.}
         \label{fig5}
   \end{figure*}

\subsection{Structure and dynamics of plume}
\label{subsec3p2}
To clarify the spatial and physical characteristics of the plume and ensure alignment with EUV observational data, Figure~\ref{fig2} (left) shows the emission integrated along \(y\) direction, rendering a cross-sectional intensity plot on the $xz$ plane. This view confirms the plume's vertical continuity and reveals alternating bright and dark emission channels, indicative of fine structuring. Figure~\ref{fig2} (right) displays the vertical velocity ($v_z$) on a $xz$ plane at a constant $y \approx 23$ Mm (see dashed line in Figure~\ref{fig3} for the reference), directly intersecting the plume region.

Three pronounced upflow regions (\(v_z\gtrsim 50\)~km\,s\(^{-1}\)) are evident in Figure~\ref{fig2} (right): (i) a strong, compact feature near \(x \sim 10\)~Mm accompanied by neighboring downflows; (ii) a narrower channel centered around \(x \sim 30\)~Mm (highlighted by a dashed oval); and (iii) a broader, intermediate-intensity region beginning near \(x \sim 45\)~Mm. In this work, we focus on the middle channel at \(x \sim 30\)~Mm for two reasons: (i) It originates at the lowest atmospheric heights (\(z \approx 3\)~Mm; dashed horizontal line)\footnote{the level used for the cut shown in Figure~\ref{fig3} (right)} and is rooted at the boundary of the plume's open magnetic funnel (positive polarity). (ii) This choice avoids boundary effects. As detailed in Figure~\ref{fig4} and Figure~\ref{fig5}, this channel coincides with the quasi-separatrix layers (QSLs) at the \textit{open–open} interface between two positive polarities, providing a direct linkage between low-lying layers and the extended corona. The associated upflow speeds reach \(v_z \approx 80\)~km\,s\(^{-1}\).

The top-down view shown in Figure~\ref{fig3} (left) illustrates the synthesized coronal emission within the EUI 174 {\AA}
filter, integrated along the vertical ($z$) direction in our model. This perspective corresponds directly to observational line-of-sight views of the Sun, highlighting prominent transient brightenings that distinctively occur away from the central plume structure. We have synthesized the emission using the same method as described in \citet{chen2021}. In this synthesized image, notable features include alternating regions of higher and lower intensity, indicative of complex magnetic and thermal dynamics in the corona.

The right panel of Figure~\ref{fig3} presents the vertical flow velocity component ($v_z$), which has been temporally averaged over approximately 18 minutes at a height of $z = 3$ Mm above the solar surface\footnote{We define the ``solar surface'' as the height corresponding to an optical depth ($\tau = 1$) in the continuum, consistent with the reference atmosphere shown in Figure~\ref{fig1}}.

A particularly striking aspect of this velocity map is the clear delineation of a robust upward flow pattern that occurs precisely at the boundary region of the plume (highlighted within the enclosed black rectangle). This pronounced upflow region underscores dynamic processes at play at plume boundaries that are responsible for the channeling of mass and energy into the corona.

For a focused investigation of the highlighted strong upflow region, we considered a vertical slice situated midway between two prominent positive magnetic polarities (Figure~\ref{fig4}). Tracking the temporal evolution along this slice reveals crucial transitions in both vertical velocity ($v_z$) and temperature ($\log T$). Notably, we find a distinct transition zone between the downward and upward flows, at a specific altitude, coinciding with a pronounced jump in the temperature from a cooler chromosphere ($\log T \approx 4.5$) to a significantly hotter corona ($\log T \approx 6.0$). This suggests that the upward-moving plasma is at coronal temperatures. In summary, our simulations highlight the role of plume boundary regions and the related low-altitude phenomena in shaping the coronal dynamics and upward mass flows.  

\subsection{What drives the upflows at the boundary of the plume?}\label{subsec3p3}
To identify the driving mechanisms behind the observed strong upflows at the plume boundaries, we analyzed the underlying magnetic field configuration. Figure~\ref{fig5}(a) presents a direct volume rendering of the squashing factor ($Q$) displayed on a natural logarithmic scale, overlaid with the magnetic field lines (black lines) from plume P and neighboring polarity P$'$. These field lines make an interface (or boundary) between two open field regions P and P$'$; hereafter referred to as \emph{OO-interface} throughout the paper. 

The squashing factor quantifies the spatial gradients in the connectivity of magnetic field lines, highlighting regions where field line mapping changes abruptly. We calculated the $Q$ values using the numerical code developed by \citet{Liu2016}, which traces magnetic field lines starting from each pixel on the bottom boundary. For a specified polarity, the code measures the separation of neighboring field lines at both their origin and endpoints. Formally, the $Q$-factor is defined by the spatial gradient of the field line mapping function, with high values indicating regions prone to rapid connectivity change, known as quasi-separatrix layers (QSLs).

Typically, high-$Q$ regions are associated with interfaces between magnetic field line configurations that are either closed-closed or open-closed, indicating distinct domains of magnetic connectivity. However, our analysis reveals an exceptionally high-$Q$ domain ($\ln$Q$ \approx$ 8) located at \emph{OO-interface}. These high-Q sheets arise due to field lines that originate from widely separated locations at the bottom boundary but geometrically converge at higher altitudes, intensifying the connectivity gradient, as illustrated in Figure~\ref{fig5}(a) and~\ref{fig5}(c).

Specifically, Figure~\ref{fig5}(b) depicts an inverted Y-shaped magnetic structure at the \emph{OO-interface}. Beyond the cusp, field lines extend upward in close proximity, enhancing the Q-factor significantly. 

Furthermore, in Figure~\ref{fig5}(c)--a top-down view of QSLs presented in Figure~\ref{fig5}(a)--  we identify that the high-$Q$ domain is highly corrugated, giving rise to multiple sheets, each characterized by a distinct $Q$ value, layered consecutively with $\ln Q$ increasing from approximately 3 to 8. Collectively, these layered sheets exhibit a remarkable filamentary structure. Moreover, the filamentary structures correlate closely with regions of enhanced emission observed in our simulations. A comparison of Figure~\ref{fig3}(a) and Figure~\ref{fig5}(c) confirms that the strongest emission aligns with these pronounced QSLs at \emph{OO-interface}, highlighting the significant influence of QSLs in shaping the plume emission structure.

\begin{figure*}[ht!]
   \centering
   \includegraphics[width=\linewidth]{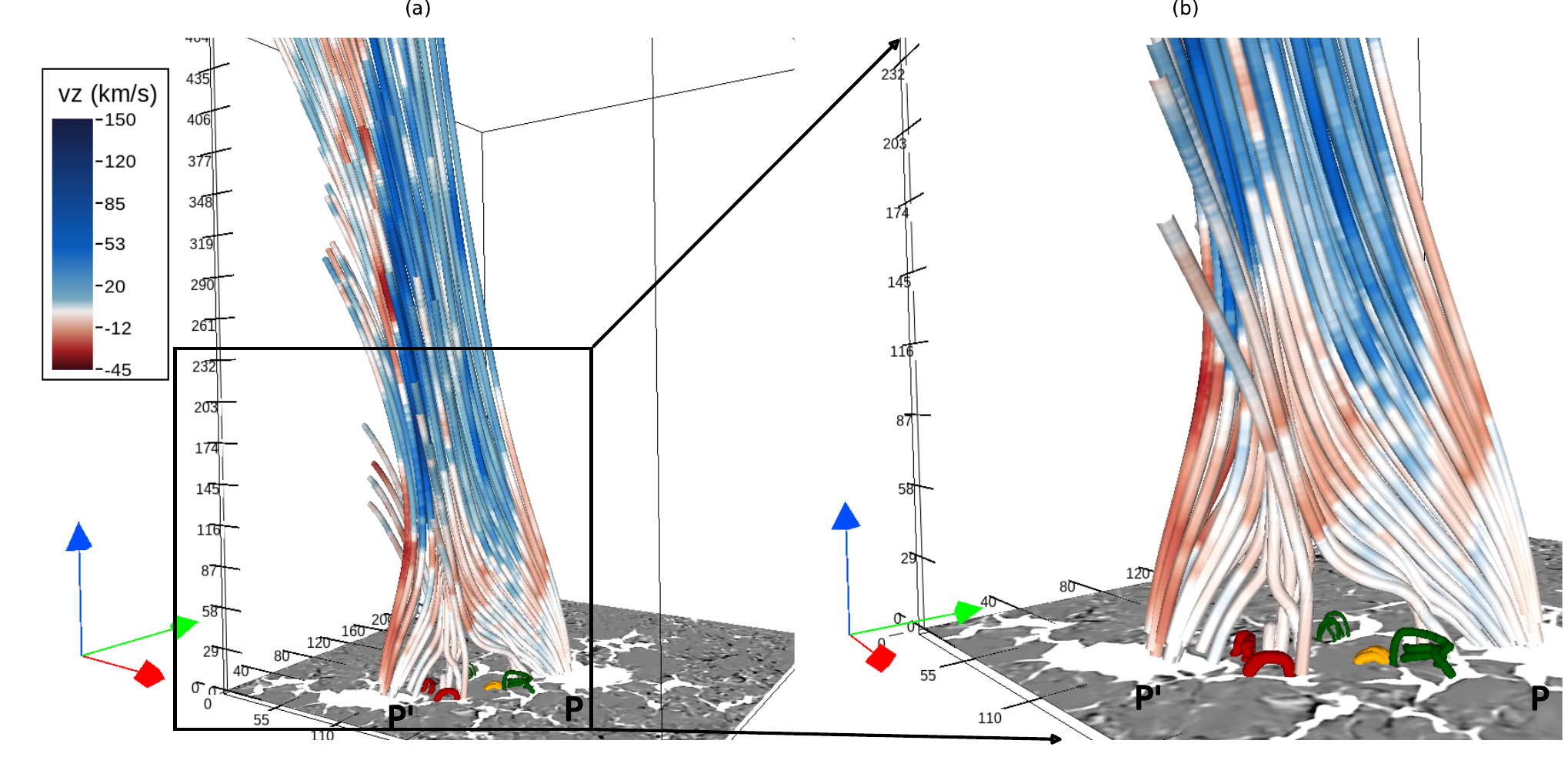}
   \caption{Snapshot depicting open magnetic field lines color-mapped by vertical velocity ($v_z$) and the closed field lines beneath them. (a) Side view of magnetic field lines, color-coded by $v_z$, emerging from the plume P and the neighboring positive-polarity P$'$. Very low-lying closed loops are highlighted in red yellow and green beneath the large-scale open field structure. (b) Enlarged view of the black-outlined region in panel (a), emphasizing the size of the underlying photospheric closed loops---the apexes reach $\approx 500~\mathrm{km}$ above the photosphere. The red, green, and blue arrows on the bottom left in each panel indicate the $x$, $y$, and $z-$axes of the 3D Cartesian coordinate system respectively. Axis labels in both panels represent grid points; along the vertical ($z$) axis, multiplying the tick values by $20~\mathrm{km}$ yields the corresponding physical heights, with $z=0$ representing the photosphere.
   See Section~\ref{subsec3p3} for details.}
         \label{fig6}
   \end{figure*}

\begin{figure*}[ht!]
   \centering
   \includegraphics[width=\linewidth]{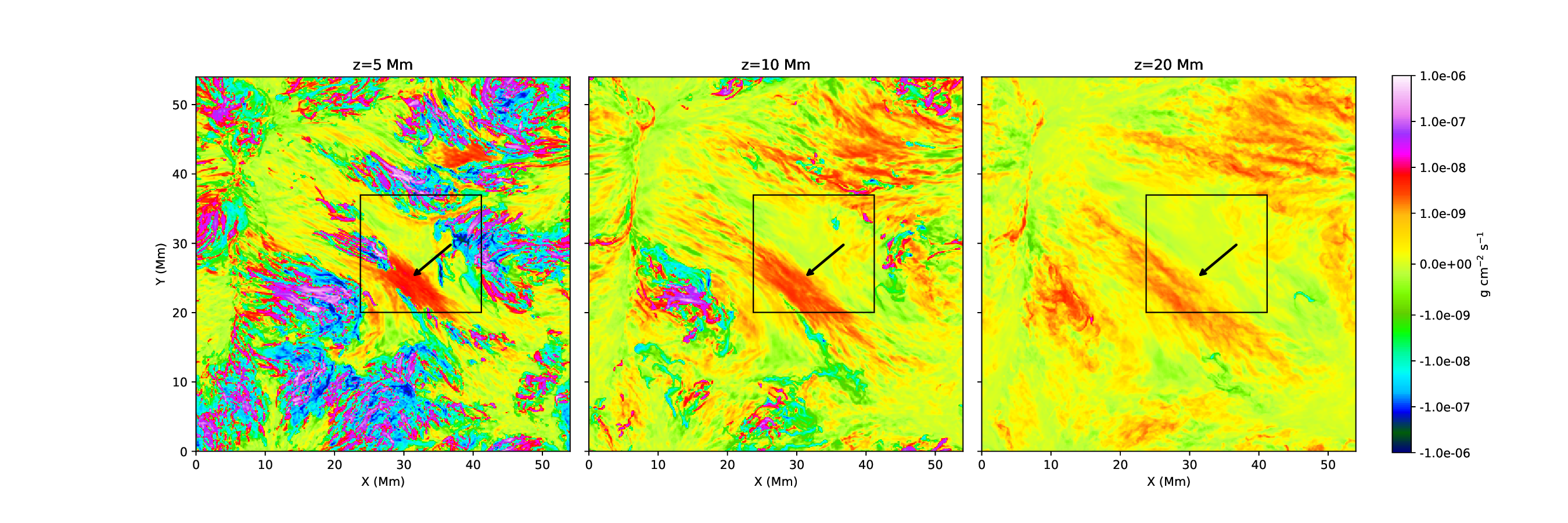}
   \caption{Time-averaged mass flux over 38 minutes, at different atmospheric heights. Three panels in the image show mass fluxes at heights $z$=5, 10, and 20\,Mm above the photosphere. The rectangle in all the images corresponds to the upflow region (same as in Figure~\ref{fig2}) and an arrow indicates the interface between two open field regions. See Section \ref{subsec3p4} for details.}
         \label{fig7}
   \end{figure*}

\subsection{Magnetic Topology and Reconnection at OO-interface.}
To test whether the intense QSL traced at the \emph{OO-interface} could simply be the halo of an embedded separatrix surface (e.g., \citealt{Pontin2016, Masson2009, Masson2017}), we carried out a dedicated topological analysis of the magnetic field both beneath and above the cusp of the inverted–Y in Figure~\ref{fig5}(b). We located magnetic null points using the trilinear method of \citet{HaynesParnell2007} applied to the full 3D data cube and then traced field lines from the neighborhoods of any identified nulls as well as from the minority-polarity patches seen below the cusp. This analysis does not reveal any null points above a height of $\approx1$~Mm anywhere under the inverted–Y (see Appendix~\ref{app:figA2} for details). Thus, any separatrices associated with nulls are confined to very low altitudes and cannot account for the high-lying QSL at the \emph{OO-interface}.

There is a population of very low-lying closed field lines rooted in the weak, opposite-polarity flux that emerges and cancels below the inverted–Y. These loops remain spatially confined beneath the strong open flux and have apex heights $\lesssim 0.5$–$0.6$~Mm. Figure~\ref{fig6}(a) shows a representative snapshot\footnote{Axis labels indicate grid indices; along the \(z\)-axis, multiply the tick values by the vertical grid spacing (20~km) to obtain physical height; for the horizontal axes (\(x\) and \(y\)), multiply the tick values by 52.734375~km. $z=0$ corresponds to the photosphere. The red, green, and blue arrows in each panel denote the $x$, $y$, and $z$ axes, respectively.} of the open field lines emanating from P and P$'$ (color-coded by $v_z$), together with a few of these closed field lines (shown in red, yellow, and green; different colors indicate different seed-point locations). Figure~\ref{fig6}(b) zooms into the region outlined in black, showing that the closed loops are restricted below $\approx 0.6$~Mm, whereas the pronounced counter-streaming flows discussed in Section~\ref{subsec3p2} originate around $2.9$~Mm above the photosphere. Taken together, the absence of any nulls above $\approx 1$~Mm and the confinement of closed loops to $\lesssim 0.58$~Mm, and the fact that the upflows arise near $\sim 3$~Mm points against null-point or interchange reconnection between open and closed fields being responsible for the observed upflows at the \emph{OO-interface}. 
We therefore conclude that the intense QSL identified at the \emph{OO-interface} is not merely the halo of a separatrix dome and that the dynamics there are more naturally explained by reconnection within (or along) the high-$Q$ sheets rather than by interchange with low-lying closed loops. 

The high-Q sheets observed in our simulation at \emph{OO-interface} closely resemble the known ``S-web'' structures described by \citet{antiochos2011}. In these structures, large $Q$-values ($>>$3) result from geometric convergence of field lines originating from spatially separate, yet same-polarity magnetic regions, creating intense localized connectivity gradients. Such S-web structures have been extensively linked to the sources of slow solar wind and are crucial for understanding coronal dynamics. 

\subsection{Can these upflows at OO-interface support the solar wind?}
\label{subsec3p4}
To assess the mass budget, we computed the vertical mass flux density, $\rho\,v_z$ (g cm$^{-2}$ s$^{-1}$) and averaged it over a 38 minutes interval across the entire simulation domain. Figure \ref{fig7} then shows this domain-averaged mass flux--each taken over the same 38 minute window--at heights of approximately 5, 10, and 20\,Mm above the photosphere. Our analysis reveals that the localized upflows, along the open-open field boundary, carry a mass flux in the range of $10^{-8}$ to $10^{-9}$\,g\,cm$^{-2}$\,s$^{-1}$ (see black arrows in Figure~\ref{fig7}).

Such fluxes are sufficient to sustain the fast solar wind, since adopting the present-day solar mass-loss rate $\dot{M}_\odot \simeq 2\times10^{-14}\,M_\odot\,\mathrm{yr}^{-1}$ \citep{Vidotto2021LRSP} and the IAU nominal solar radius $R_\odot$ \citep{Prsa2016AJ} yields a global mean surface mass flux of $\dot{M}_\odot/(4\pi R_\odot^{2}) \approx 1.6\times10^{-11}\,\mathrm{g}\,\mathrm{cm}^{-2}\,\mathrm{s}^{-1}$. The black rectangle in the $z=5\,$~Mm map marks the field of view of Figures~\ref{fig4}, \ref{fig5}, and \ref{fig6}, where we examined the plume core and its boundary upflows in greater detail. The positive (upward) mass flux consistently occurs along the QSLs between two open‐field regions--the plume boundary--where strong plasma upflows are channeled along open field lines. 
At the $z=5$~Mm slice, regions away from the open–open boundary exhibit both positive (upward) and strong negative (downward) mass fluxes, with the latter of order $-10^{-8}$ to $-10^{-7}\,\mathrm{g\,cm^{-2}\,s^{-1}}$, indicative of return flows. These negative fluxes decay with height--by $z=20\,\mathrm{Mm}$ they fall below $\sim10^{-9}\,\mathrm{g\,cm^{-2}\,s^{-1}}$, whereas along the open–open boundary the upflow signature remains a continuous positive mass flux across all sampled heights. This robustness suggests that the open-open boundary upflows emerging from the coronal base of plumes are well capable of feeding the nascent solar wind.
 
\section{Discussion and Conclusions}

In this study, we employed high-resolution magnetohydrodynamic (MHD) simulations to examine the structure and evolution of a plume within a coronal-hole environment. Our results provide insights into the nature of plume emission, the dynamics at magnetic-field boundaries, and the origins of the persistent solar wind, offering both support and challenges to prevailing theoretical models.

Extreme Ultraviolet (EUV) observations have consistently shown that the bases of coronal-hole plumes are populated by highly dynamic, small-scale jet-like features, variously termed jetlets or plumelets \citep{raouafi2014}. These features are thought to arise from localized magnetic reconnection between emerging small-scale bipoles and pre-existing ambient open magnetic fields, leading to energy release and plasma ejection. Such reconnection-driven jets are believed to contribute to the mass supply to the corona and to the solar wind.

In contrast, our simulation does not reproduce jetlets or similar features at the base of the plume in synthesized EUV emission. Instead, we observe transient brightenings located away from the plume region, linked to reconnection events external to the core open-field structure. When viewed along certain lines of sight, these brightenings can project onto the plume, creating the illusion of localized activity within it. This highlights the critical role of viewing geometry in interpreting EUV observations and emphasizes the need for multi-perspective and stereoscopic imaging, such as that provided by STEREO and Solar Orbiter, to disentangle line-of-sight effects.

The simulated plume exhibits a well-defined filamentary morphology, comprising alternating bright and dark channels in synthetic EUI 174~\AA\ emission, consistent with the filamentary structures reported by \citet{uritsky2021}. Although our simulation achieves high spatial resolution, it still falls short of producing the sub-arcsecond-scale jets (a few hundred kilometers wide) identified by \citet{chitta2023} using Solar Orbiter's Extreme Ultraviolet Imager (EUI). 
Though the plume footpoint size ($\sim$2--4~Mm) and flux ($\sim 10^{19}$~Mx) are comparable to the observations \citet{DeForest1997}, the lack of small-scale jets may be a result of the high magnetic field strength at the plume base, which slightly exceeds the observed kilogauss values \citep{Tsuneta2008} in addition to the suppressed convection within the flux tube.

While previous studies have emphasized emerging bipoles and null-point reconnection \citep{Kumar2022}, our simulation shows that fine-structured plume emission does not rely solely on reconnection at a null point or on bipole emergence, suggesting that more generic mechanisms may play a key role. 

A central finding of our study is that the \emph{OO-interfaces} play a key role in structuring plumes and channeling plasma flows. Along these interfaces, the initially coherent open magnetic flux tube becomes fragmented by convective motions, leading to lateral displacement and shearing of magnetic polarities. The resulting configurations, reminiscent of inverted-Y or $\lambda$ shapes, guide plasma dynamics in ways not captured by models focused solely on null-point reconnection.

One clear feature captured in our simulation is the transition from cool downflows to hot upflows in the low corona. This thermodynamic structure is consistent with spectroscopic observations by SUMER and EIS \citep{wilhelm1995, tian2010}, which report localized hot outflows within coronal holes. These upflows are concentrated within finely structured quasi-separatrix layers (QSLs) that form at the \emph{OO-interface} between the plume and neighboring open field. Such QSLs may act as localized ``nozzles'' that channel focused plasma flows, analogous to solar-wind corridors described in the S-web model of \citet{antiochos2011}. A similar configuration has been invoked to explain persistent diffuse coronal structures located between like-polarity magnetic-flux concentrations \citep[][]{milanovic2023}.

The magnetic field lines forming these QSLs are non-parallel as they extend upward and may favor component reconnection driven by the mutual misalignment of field lines belonging to two open flux tubes. While the concept of component reconnection in open-field regions is well established \citep[e.g.,][]{Priest2002, Matthaeus1999, Leamon2000, Dmitruk2002, Servidio2010, Rappazzo2012}, our results demonstrate how plume-scale QSLs provide a specific topological context that can organize reconnection-driven upflows into fine-structured channels.

Moreover, the absence of emerging opposite polarities within the plume challenges models that rely on internal bipole emergence for plume dynamics and fine-structure \citep{raouafi2014, Kumar2022}. This finding is consistent with the results of \citet{Avallone2018}, who showed that plume formation and long-term visibility are primarily regulated by the concentration of unipolar magnetic flux, without requiring continual flux emergence or cancellation. Their analysis suggests that sufficient flux convergence alone can sustain plume heating and emission, supporting our view that stable plume bodies can persist in purely unipolar magnetic environments. However, in our simulation the upflows do not trace the QSL layers at higher altitudes, as the plume field expands, the QSLs at its boundary diverge from upflows (see Appendix \ref{flow_logQ}). The physical mechanism that initiates the upflows along QSLs therefore remains unresolved and is an open question.

To assess whether the intense QSL at the \emph{OO-interface could simply be the halo of a separatrix dome beneath the inverted-Y, we conducted a dedicated search for magnetic nulls using a trilinear approach \citep{HaynesParnell2007}. We do not find nulls at the heights where the counter-streaming flows or hot upflows originate; any associated closed loops are confined to very low altitudes. This height separation argues against null-point or interchange reconnection as the driver of the reported upflows and instead supports a QSL-mediated process.}

In summary, our results show that plume fine structure and upflows can be sustained solely by the dynamics at the \emph{OO-interface}, without requiring flux emergence or null-point reconnection. The narrow lanes of upflow originating from such regions exhibit consistent upward mass fluxes capable of feeding the nascent solar wind, reinforcing their significance in heliospheric mass supply. This broader mechanism---rooted in open–open magnetic interactions---adds to our evolving understanding of solar-wind origins and provides a framework for future observational and theoretical efforts aimed at distinguishing the relative roles of different reconnection processes.

\begin{acknowledgements}
     We thank the anonymous reviewer for insightful comments that improved the accuracy and clarity of this paper. K.B. and L.P.C. gratefully acknowledge funding by the European Union (ERC, ORIGIN, 101039844). Views and opinions expressed are however those of the author(s) only and do not necessarily reflect those of the European Union or the European Research Council. Neither the European Union nor the granting authority can be held responsible for them. 
     Y.C. acknowledges funding provided by the Alexander von Humboldt Foundation. The work of Y.C. and D. Przybylski was funded by the Federal Ministry for Economic Affairs and Climate Action (BMWK) through the German Space Agency at DLR based on a decision of the German Bundestag (Funding code: 50OU2201).
     We gratefully acknowledge the computational resources provided by the Cobra \& Raven supercomputer systems of the Max Planck Computing and Data Facility (MPCDF) in Garching, Germany. 
     We wish to acknowledge the visualization software VAPOR \url{(www.vapor.ucar.edu)}, for generating direct volume renderings and field lines plots.
\end{acknowledgements}

\newpage

\appendix

\section{Method for Inserting a Magnetic Plume and Subsequent Staged Coronal Extension in the MURaM Simulation Domain}\label{ap1}
To enable controlled studies of flux emergence and magnetic structure evolution in the solar atmosphere, we developed a method to embed a localized, open magnetic flux tube (plume) into a MURaM simulation snapshot. This approach ensures that, initially, the inserted structure (1) has an exponential vertical decay in peak field strength; (2) preserves constant total magnetic flux with height by appropriately expanding its cross-section; (3) is realized as a potential (divergence-free and current-free) magnetic field in each horizontal plane; (4) is spatially confined to prevent field leakage; and (5) minimally disturbs the preexisting ambient field, altering only the region actually occupied by the plume. 

Prior to adding the chromosphere and coronal extension, the computational domain from MURaM snapshot has $1050 \times 1024 \times 1024$ grid points along $(z,x,y)$, with $z=0$ at the convection-zone base and $z=1036$ and $z=1049$ marking the photosphere and top boundary, respectively.

A three‐dimensional plume is constructed and merged with the background magnetic data in several well-defined stages, as follows.

\subsection{Flux-Tube Geometry, Vertical Profile, and Flux Conservation}

The plume's axis is aligned with the vertical ($z$) direction and centered horizontally at $(x_c, y_c)$. At each discrete height $i$, the tube's cross-section is prescribed as a two-dimensional Gaussian in $(x, y)$:
\begin{equation}
  B_z(x, y, i) = B_0(i)\exp\left(-\frac{(x - x_c)^2 + (y - y_c)^2}{2\sigma(i)^2}\right),
\end{equation}
where $B_0(i)$ is the peak field at height $i$ and $\sigma(i)$ is the Gaussian half-width.

The vertical ($z$-dependent) magnetic field strength $B_0(i)$ within the tube is designed to decrease exponentially with height, reflecting the natural stratification of solar atmospheric pressure and density. It is anchored at $B_{\mathrm{base}} = 20{,}000\,\mathrm{G}$ at $i=0$, yielding a scale height
\begin{equation}
  H = -\frac{1050}{\ln(1500/10000)} \approx 553.47,
\end{equation}
so that
\begin{equation}
  B_0(i) = 20{,}000\exp\left(-\frac{i}{H}\right).
  \label{eq:Bz_exponential_app}
\end{equation}

This reflects the natural stratification of solar atmospheric pressure and density. At the photosphere ($i=1036$), this gives $B_0(1036) \approx 3.07\times 10^3\,\mathrm{G}$, a field strength that is roughly a factor 1.5 to 2 larger than in the observed pores \citep[][]{suetterlin1998}. Insertion of a flux tube with such a strong magnetic field strength was required to counter the effects of convective shredding of the pore. Field strength in the simulation domain was reduced by a factor of 1.8 in the latter stages of evolution to ensure comparability with observations. 

To maintain a constant total vertical magnetic flux along the length of the tube, the cross-sectional width of the tube $\sigma(i)$ is inversely proportional to the square root of the magnetic field strength $B_0(i)$. This ensures the integrated magnetic flux across any horizontal cross-section remains uniform throughout the domain:
\begin{equation}
  \Phi(i) = B_0(i)\,\pi\,\sigma(i)^2 = \text{constant}.
\end{equation}

Given an initial half-width $\sigma_0$ at $i=0$, the width at height $i$ is set as
\begin{equation}
  \sigma(i) = \sigma_0 \sqrt{\frac{B_0(1000)}{B_0(i)}},
\end{equation}
where $\sigma_0$ is chosen so that the full-width at half-maximum (FWHM) near the photosphere is approximately $2\,\mathrm{Mm}$:
\begin{equation}
  \sigma_0 = \frac{\mathrm{FWHM}}{2\sqrt{2\ln 2}}.
\end{equation}

\subsection{Potential-Field Extrapolation and Field-Line Structure}

At each height, the imposed vertical field $B_z(x, y, i)$ serves as the boundary condition for a potential (current-free) magnetic field in the horizontal plane:
\begin{equation}
  \nabla\times\mathbf{B} = \mathbf{0}, \quad \nabla\cdot\mathbf{B} = 0.
\end{equation}
The horizontal field components $(B_x, B_y)$ are obtained by solving Laplace’s equation for the scalar potential (using Fast Fourier Transform (FFT) approach), yielding a smooth, divergence-free, and curl-free magnetic structure whose vertical profile matches the Gaussian prescription.

\subsection{Finite Horizontal Confinement via Masking}

To confine the plume and prevent artificial field leakage, a circular mask of radius
\begin{equation}
  r_{\mathrm{mask}}(i) = 5\,\sigma(i)
\end{equation}
is applied at each height $i$. The magnetic field is set to zero outside this mask:
\begin{equation}
  \mathbf{B}_{\mathrm{plume}}(x, y, i) =
  \begin{cases}
    (B_x, B_y, B_z)(x, y, i), & \text{if } (x - x_c)^2 + (y - y_c)^2 \le [5\,\sigma(i)]^2, \\[6pt]
    (0, 0, 0), & \text{otherwise.}
  \end{cases}
\end{equation}

\subsection{Merging with the Background and Minimal Disturbance}

The existing three-dimensional magnetic data cubes for $B_x$, $B_y$, and $B_z$ serve as the ambient background:
\begin{equation}
  [B_z^{(\mathrm{bg})}, B_x^{(\mathrm{bg})}, B_y^{(\mathrm{bg})}](x, y, i).
\end{equation}

A complementary mask $M_{\mathrm{bg}}(x, y; i)$ is used (unity outside $r_{\mathrm{mask}}(i)$, zero inside), such that the final merged field is
\begin{equation}
  B_\alpha^{(\mathrm{new})}(x, y, i) =
    B_\alpha^{(\mathrm{bg})}(x, y, i)\,M_{\mathrm{bg}}(x, y; i)
    + B_\alpha^{(\mathrm{plume})}(x, y, i), \quad \alpha \in \{x, y, z\}.
\end{equation}

The plume field replaces the original field only within its defined footprint, ensuring a minimal and localized disturbance. The resulting data set contains a physically motivated, divergence-free magnetic tube embedded into the MURaM snapshot, serving as initial condition for dynamic MHD evolution.

\subsection{Subsequent  coronal extension and relaxation}
Following insertion of the flux tube, the setup was evolved for additional several solar hours without a corona, with the top of the simulation domain located at approximately $500$\,km above the photosphere. This phase allowed the flux tube to interact dynamically with the ambient field and to achieve a relaxed state of photospheric magnetoconvection. We then raised the top boundary to $6$\,Mm by performing a potential-field extrapolation of the magnetic field from the previous top boundary and populated the newly added layers with a hydrostatic, isothermal atmosphere initialized from the horizontal mean pressure and temperature at the previous upper boundary. This sequential extension of the upper boundary is mainly to mitigate the very small computational timesteps that arise from very high Alfv\'en speeds in the corona. After a transition region had formed and the temperature reached a few times $10^{5}$\,K, we expanded the top boundary by another $24$\,Mm and continued with the field-aligned Spitzer heat conduction and optically thin radiative losses in the upper atmosphere as described in \citet{rempel2017}, consistent with recent quiet-Sun coronal applications \citep{Chen2025}.

\section{Height Dependence of QSLs and Time-Averaged Vertical Velocity in the Simulation Domain}
\label{flow_logQ}
\begin{figure*}[ht!]
    \centering
    \includegraphics[width=1\linewidth, trim={0.5cm 1cm 0.5cm 0},clip]{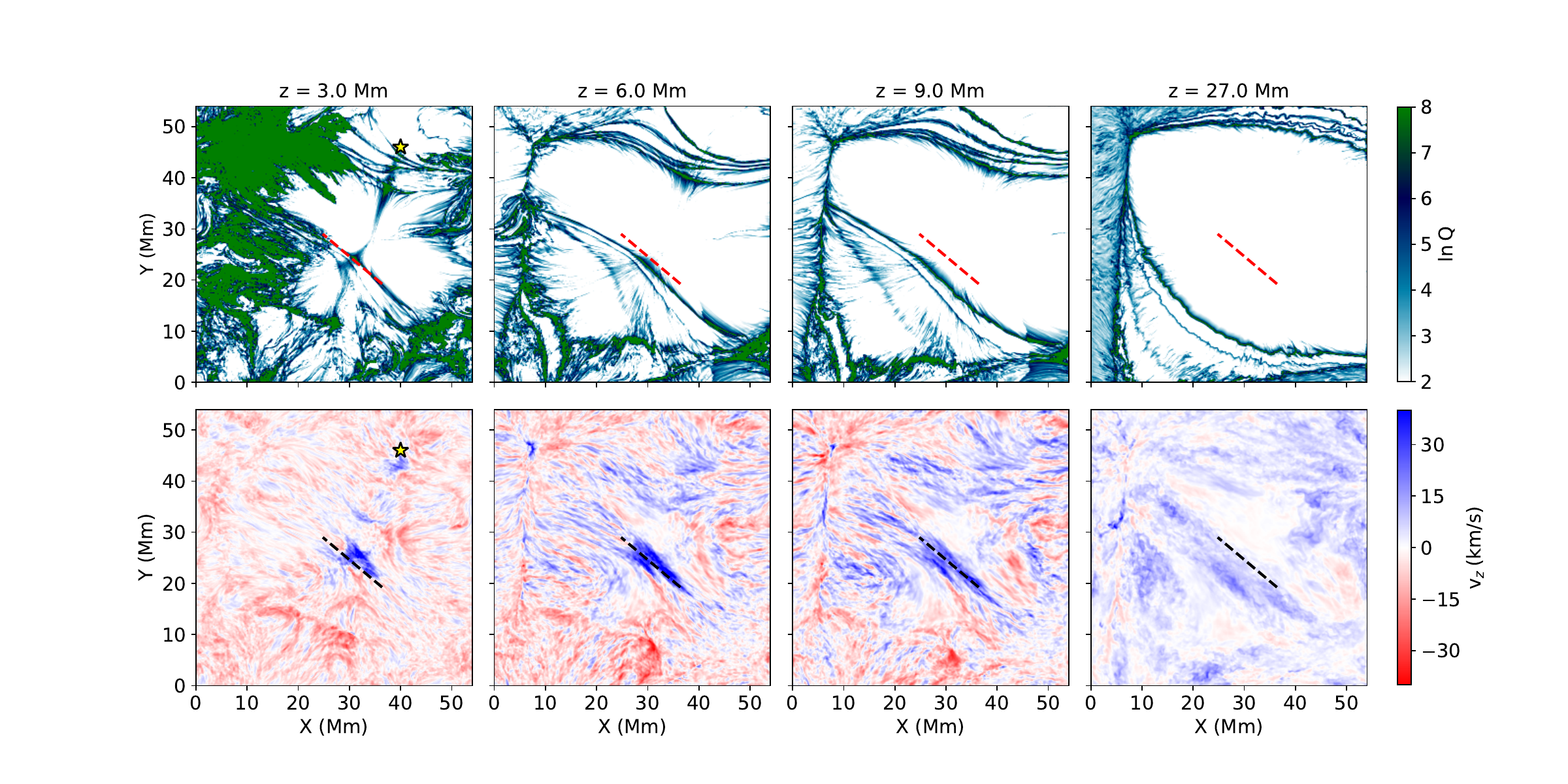}
    \caption{Upflow structure associated with $\ln Q$. The top row shows maps of $\ln Q$ at $z=3$, 6, 9 and 27\,Mm, while the bottom row shows the corresponding vertical velocity $v_z$, time-averaged over 18\,minutes, at the same heights and over the full horizontal extent. The red (top panels) and black (bottom panels) dashed lines trace a reference path at \emph{OO-interface}, marking the plume boundary. The star symbol marks the other upflow region that also originates at \emph{OO-interface} and coincides with high-$Q$ sheet. Although the high-$Q$ layers diverge from this boundary with increasing height, the upflows remain confined to it.}
    \label{fig:appendix1}
\end{figure*}

Figure~\ref{fig:appendix1} presents the vertical structure of the upflow region in relation to the squashing factor, $Q$, across four heights in our simulation domain. In the top row, we plot maps of $\ln Q$ at heights of $z = 3$, 6, 9, and 27\,Mm, capturing the evolving geometry of the magnetic connectivity gradient. The bottom row shows the corresponding vertical velocity, $v_{z}$, at the same heights, each map representing a 18 minutes temporal average to highlight persistent upflow patterns. 
A dashed line in each panel traces a reference path at \emph{OO-interface} and delineates the boundary of the plume. In the $\ln Q$ maps, these high‐$Q$ layers emanate at the open field interface but fan outward with increasing altitude, indicating that the magnetic connectivity gradient is affected by the field expansion. By contrast, the upflow pattern remains closely collimated along the same reference path at all heights, demonstrating that the strongest vertical motions are confined to the original plume boundary even as the magnetic topology opens up around them. Other upflow region (not discussed above) is marked by a star and originates at the other \emph{OO-interface} on the right hand side of the plume along diagonal direction (perpendicular to dashed line). This upflow also causes the upward (positive) mass-flux (cf. Figure~\ref{fig7} at z=10 and 20\,Mm) and is associated with the QSL reconnection.

\section{Null points distribution with height}
\label{app:figA2}
\begin{figure*}[ht!]
    \centering
    \includegraphics[width=1\linewidth]{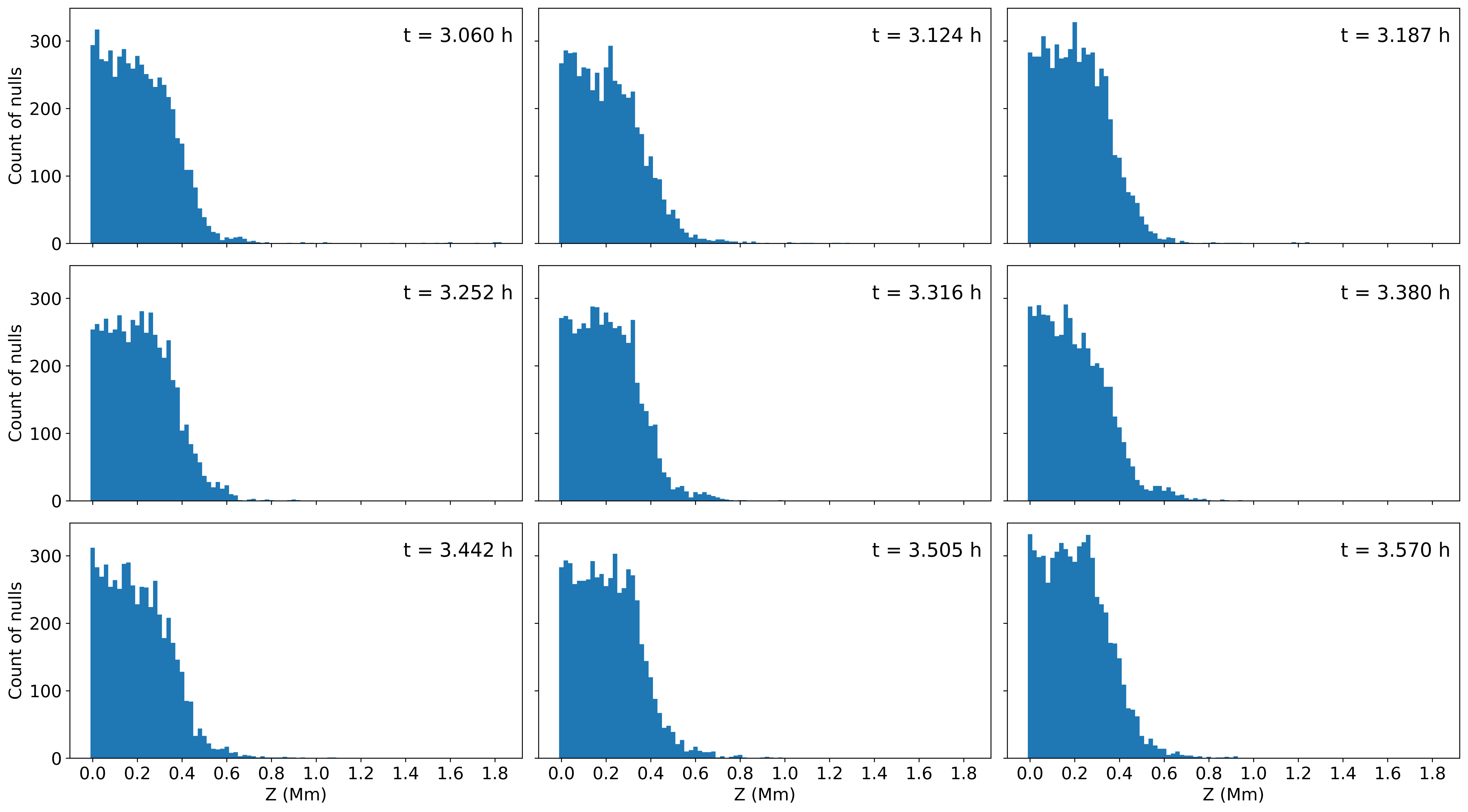}
    \caption{Height distribution of magnetic nulls in the P--P$'$ region. Histograms show the count of magnetic nulls as a function of height $z$ (Mm) for ten simulation snapshots, at $\Delta t=3.8$ minutes intervals, spanning $t=3.060$--$3.570$~h. Statistics are computed inside a fixed sub-volume between plume P and the neighboring positive polarity P$'$ (extent: 7.91~Mm$\times$9.5~Mm$\times$10~Mm along $x$, $y$, and $z-$directions respectively; exact bounds given in Appendix~\ref{app:figA2}). All panels share the same axes and binning. The distribution is strongly weighted toward low altitudes, with very few nulls above $\sim 0.6$~Mm, indicating a dense carpet of topological features well below the $\sim 1$~Mm height.}
    \label{fig:A2}
\end{figure*}

To examine the magnetic topology in between plume P and the neighboring positive polarity P$'$, we selected a fixed Cartesian sub-volume that encompasses the region where Figures~2 and~4 reveal a sharp thermal/dynamic transition at $z \approx 3$~Mm. The sub-volume's geometric extent is 7.91~Mm$\times$9.5~Mm$\times$10~Mm along $x$, $y$, and $z-$directions respectively; for reproducibility, the precise
limits are $[x_{\min},x_{\max}] \times [y_{\min},y_{\max}] \times [z_{\min},z_{\max}] =
[27.94,35.85] \times [20.03,29.53] \times [0,10]$~Mm,
which remain fixed across all snapshots.
Within this volume, magnetic nulls were found in each stored snapshot, and their heights were binned uniformly in $z$ to produce the histograms shown in Figure~\ref{fig:A2}. The snapshots are spaced by $\Delta t=3.8$~minutes and cover $t=3.060$--$3.570$~h, approximately the time interval analyzed in the main text. All panels use identical vertical axes (counts of nulls) and horizontal axes ($z$ in Mm) with the same binning; the sub-volume and cadence are held fixed in time, so variations reflect genuine temporal evolution rather than changes in sampling.
Across the entire sequence, the null population remains concentrated at very low heights, with a steep drop-off above $\sim 0.6$-$1$~Mm. This persistent low-lying topology is consistent with frequent small-scale reconnection sites in the near-surface magnetic carpet. The scarcity of nulls above 0.6$-$1~Mm height suggests that the pronounced transition from cool downflows to hot upflows reported in Figures~\ref{fig2} and~\ref{fig4} is neither directly associated to any null point topology at $z \approx 3$~Mm nor rooted in topologically complex magnetic carpet field.

\bibliography{References_aastex}{}

@ARTICLE{uritsky2021,
       author = {{Uritsky}, V.~M. and {DeForest}, C.~E. and {Karpen}, J.~T. and {DeVore}, C.~R. and {Kumar}, P. and {Raouafi}, N.~E. and {Wyper}, P.~F.},
        title = "{Plumelets: Dynamic Filamentary Structures in Solar Coronal Plumes}",
      journal = {\apj},
         year = 2021,
        month = jan,
       volume = {907},
       number = {1},
          eid = {1},
        pages = {1},
          doi = {10.3847/1538-4357/abd186},
archivePrefix = {arXiv},
       eprint = {2012.05728},
 primaryClass = {astro-ph.SR},
       adsurl = {https://ui.adsabs.harvard.edu/abs/2021ApJ...907....1U}
}

@ARTICLE{tian2011,
       author = {{Tian}, Hui and {McIntosh}, Scott W. and {Habbal}, Shadia Rifai and {He}, Jiansen},
        title = "{Observation of High-speed Outflow on Plume-like Structures of the Quiet Sun and Coronal Holes with Solar Dynamics Observatory/Atmospheric Imaging Assembly}",
      journal = {\apj},
         year = 2011,
        month = aug,
       volume = {736},
       number = {2},
          eid = {130},
        pages = {130},
          doi = {10.1088/0004-637X/736/2/130},
archivePrefix = {arXiv},
       eprint = {1105.3119},
 primaryClass = {astro-ph.SR},
       adsurl = {https://ui.adsabs.harvard.edu/abs/2011ApJ...736..130T}
}

@INPROCEEDINGS{tian2010,
       author = {{Tian}, H. and {Tu}, C. -Y. and {Marsch}, E. and {He}, J. -S. and {Zhou}, C. and {Zhao}, L.},
        title = "{Upflows in the upper transition region of the quiet Sun}",
    booktitle = {Twelfth International Solar Wind Conference},
         year = 2010,
       editor = {{Maksimovic}, M. and {Issautier}, K. and {Meyer-Vernet}, N. and {Moncuquet}, M. and {Pantellini}, F.},
       series = {American Institute of Physics Conference Series},
       volume = {1216},
        month = mar,
    publisher = {AIP},
        pages = {36-39},
          doi = {10.1063/1.3395877},
archivePrefix = {arXiv},
       eprint = {0911.1833},
 primaryClass = {astro-ph.SR},
       adsurl = {https://ui.adsabs.harvard.edu/abs/2010AIPC.1216...36T}
}

@ARTICLE{antiochos2011,
       author = {{Antiochos}, S.~K. and {Miki{\'c}}, Z. and {Titov}, V.~S. and {Lionello}, R. and {Linker}, J.~A.},
        title = "{A Model for the Sources of the Slow Solar Wind}",
      journal = {\apj},
         year = 2011,
        month = apr,
       volume = {731},
       number = {2},
          eid = {112},
        pages = {112},
          doi = {10.1088/0004-637X/731/2/112},
archivePrefix = {arXiv},
       eprint = {1102.3704},
 primaryClass = {astro-ph.SR},
       adsurl = {https://ui.adsabs.harvard.edu/abs/2011ApJ...731..112A}
}

@ARTICLE{cranmer2002,
       author = {{Cranmer}, Steven R.},
        title = "{Coronal Holes and the High-Speed Solar Wind}",
      journal = {\ssr},
         year = 2002,
        month = aug,
       volume = {101},
       number = {3},
        pages = {229-294},
          doi = {10.1023/A:1020840004535},
       adsurl = {https://ui.adsabs.harvard.edu/abs/2002SSRv..101..229C}
}

@ARTICLE{Kumar2022,
       author = {{Kumar}, Pankaj and {Karpen}, Judith T. and {Uritsky}, Vadim M. and {Deforest}, Craig E. and {Raouafi}, Nour E. and {Richard DeVore}, C.},
        title = "{Quasi-periodic Energy Release and Jets at the Base of Solar Coronal Plumes}",
      journal = {\apj},
         year = 2022,
        month = jul,
       volume = {933},
       number = {1},
          eid = {21},
        pages = {21},
          doi = {10.3847/1538-4357/ac6c24},
archivePrefix = {arXiv},
       eprint = {2204.13871},
 primaryClass = {astro-ph.SR},
       adsurl = {https://ui.adsabs.harvard.edu/abs/2022ApJ...933...21K}
}

@ARTICLE{raouafi2014,
       author = {{Raouafi}, N. -E. and {Stenborg}, G.},
        title = "{Role of Transients in the Sustainability of Solar Coronal Plumes}",
      journal = {\apj},
         year = 2014,
        month = jun,
       volume = {787},
       number = {2},
          eid = {118},
        pages = {118},
          doi = {10.1088/0004-637X/787/2/118},
       adsurl = {https://ui.adsabs.harvard.edu/abs/2014ApJ...787..118R}
}

@ARTICLE{wang2009,
       author = {{Wang}, Y. -M.},
        title = "{Coronal Holes and Open Magnetic Flux}",
      journal = {\ssr},
         year = 2009,
        month = apr,
       volume = {144},
       number = {1-4},
        pages = {383-399},
          doi = {10.1007/s11214-008-9434-0},
       adsurl = {https://ui.adsabs.harvard.edu/abs/2009SSRv..144..383W}
}

@ARTICLE{tu2005,
       author = {{Tu}, Chuan-Yi and {Zhou}, Cheng and {Marsch}, Eckart and {Xia}, Li-Dong and {Zhao}, Liang and {Wang}, Jing-Xiu and {Wilhelm}, Klaus},
        title = "{Solar Wind Origin in Coronal Funnels}",
      journal = {Science},
         year = 2005,
        month = apr,
       volume = {308},
       number = {5721},
        pages = {519-523},
          doi = {10.1126/science.1109447},
       adsurl = {https://ui.adsabs.harvard.edu/abs/2005Sci...308..519T}
}

@ARTICLE{wilhelm1995,
       author = {{Wilhelm}, K. and {Curdt}, W. and {Marsch}, E. and {Sch{\"u}hle}, U. and {Lemaire}, P. and {Gabriel}, A. and {Vial}, J. -C. and {Grewing}, M. and {Huber}, M.~C.~E. and {Jordan}, S.~D. and {Poland}, A.~I. and {Thomas}, R.~J. and {K{\"u}hne}, M. and {Timothy}, J.~G. and {Hassler}, D.~M. and {Siegmund}, O.~H.~W.},
        title = "{SUMER - Solar Ultraviolet Measurements of Emitted Radiation}",
      journal = {\solphys},
         year = 1995,
        month = dec,
       volume = {162},
       number = {1-2},
        pages = {189-231},
          doi = {10.1007/BF00733430},
       adsurl = {https://ui.adsabs.harvard.edu/abs/1995SoPh..162..189W}
}

@ARTICLE{fisk2005,
       author = {{Fisk}, L.~A.},
        title = "{The Open Magnetic Flux of the Sun. I. Transport by Reconnections with Coronal Loops}",
      journal = {\apj},
         year = 2005,
        month = jun,
       volume = {626},
       number = {1},
        pages = {563-573},
          doi = {10.1086/429957},
       adsurl = {https://ui.adsabs.harvard.edu/abs/2005ApJ...626..563F}
}

@ARTICLE{priest1995,
       author = {{Priest}, E.~R. and {D{\'e}moulin}, P.},
        title = "{Three-dimensional magnetic reconnection without null points. 1. Basic theory of magnetic flipping}",
      journal = {\jgr},
         year = 1995,
        month = dec,
       volume = {100},
       number = {A12},
        pages = {23443-23464},
          doi = {10.1029/95JA02740},
       adsurl = {https://ui.adsabs.harvard.edu/abs/1995JGR...10023443P}
}

@ARTICLE{demoulin1996,
       author = {{Demoulin}, P. and {Henoux}, J.~C. and {Priest}, E.~R. and {Mandrini}, C.~H.},
        title = "{Quasi-Separatrix layers in solar flares. I. Method.}",
      journal = {\aap},
         year = 1996,
        month = apr,
       volume = {308},
        pages = {643-655},
       adsurl = {https://ui.adsabs.harvard.edu/abs/1996A&A...308..643D}
}

@ARTICLE{voegler2005,
       author = {{V{\"o}gler}, A. and {Shelyag}, S. and {Sch{\"u}ssler}, M. and {Cattaneo}, F. and {Emonet}, T. and {Linde}, T.},
        title = "{Simulations of magneto-convection in the solar photosphere.  Equations, methods, and results of the MURaM code}",
      journal = {\aap},
         year = 2005,
        month = jan,
       volume = {429},
        pages = {335-351},
          doi = {10.1051/0004-6361:20041507},
       adsurl = {https://ui.adsabs.harvard.edu/abs/2005A&A...429..335V}
}

@ARTICLE{rempel2017,
       author = {{Rempel}, M.},
        title = "{Extension of the MURaM Radiative MHD Code for Coronal Simulations}",
      journal = {\apj},
         year = 2017,
        month = jan,
       volume = {834},
       number = {1},
          eid = {10},
        pages = {10},
          doi = {10.3847/1538-4357/834/1/10},
archivePrefix = {arXiv},
       eprint = {1609.09818},
 primaryClass = {astro-ph.SR},
       adsurl = {https://ui.adsabs.harvard.edu/abs/2017ApJ...834...10R}
}

@BOOK{spitzer1962,
       author = {{Spitzer}, L.},
        title = "{Physics of Fully Ionized Gases}",
         year = 1962,
       adsurl = {https://ui.adsabs.harvard.edu/abs/1962pfig.book.....S}
}

@ARTICLE{vogler2007,
       author = {{V{\"o}gler}, A. and {Sch{\"u}ssler}, M.},
        title = "{A solar surface dynamo}",
      journal = {\aap},
         year = 2007,
        month = apr,
       volume = {465},
       number = {3},
        pages = {L43-L46},
          doi = {10.1051/0004-6361:20077253},
archivePrefix = {arXiv},
       eprint = {astro-ph/0702681},
 primaryClass = {astro-ph},
       adsurl = {https://ui.adsabs.harvard.edu/abs/2007A&A...465L..43V}
}

@ARTICLE{cranmer2009,
       author = {{Cranmer}, Steven R.},
        title = "{Coronal Holes}",
      journal = {Living Reviews in Solar Physics},
         year = 2009,
        month = dec,
       volume = {6},
       number = {1},
          eid = {3},
        pages = {3},
          doi = {10.12942/lrsp-2009-3},
archivePrefix = {arXiv},
       eprint = {0909.2847},
 primaryClass = {astro-ph.SR},
       adsurl = {https://ui.adsabs.harvard.edu/abs/2009LRSP....6....3C}
}

@ARTICLE{chitta2023,
       author = {{Chitta}, L.~P. and {Zhukov}, A.~N. and {Berghmans}, D. and {Peter}, H. and {Parenti}, S. and {Mandal}, S. and {Aznar Cuadrado}, R. and {Sch{\"u}hle}, U. and {Teriaca}, L. and {Auch{\`e}re}, F. and {Barczynski}, K. and {Buchlin}, {\'E}. and {Harra}, L. and {Kraaikamp}, E. and {Long}, D.~M. and {Rodriguez}, L. and {Schwanitz}, C. and {Smith}, P.~J. and {Verbeeck}, C. and {Seaton}, D.~B.},
        title = "{Picoflare jets power the solar wind emerging from a coronal hole on the Sun}",
      journal = {Science},
         year = 2023,
        month = aug,
       volume = {381},
       number = {6660},
        pages = {867-872},
          doi = {10.1126/science.ade5801},
archivePrefix = {arXiv},
       eprint = {2308.13044},
 primaryClass = {astro-ph.SR},
       adsurl = {https://ui.adsabs.harvard.edu/abs/2023Sci...381..867C}
}

@ARTICLE{gabriel2009,
       author = {{Gabriel}, A. and {Bely-Dubau}, F. and {Tison}, E. and {Wilhelm}, K.},
        title = "{The Structure and Origin of Solar Plumes: Network Plumes}",
      journal = {\apj},
         year = 2009,
        month = jul,
       volume = {700},
       number = {1},
        pages = {551-558},
          doi = {10.1088/0004-637X/700/1/551},
       adsurl = {https://ui.adsabs.harvard.edu/abs/2009ApJ...700..551G}
}

@ARTICLE{Liu2016,
       author = {{Liu}, Rui and {Kliem}, Bernhard and {Titov}, Viacheslav S. and {Chen}, Jun and {Wang}, Yuming and {Wang}, Haimin and {Liu}, Chang and {Xu}, Yan and {Wiegelmann}, Thomas},
        title = "{Structure, Stability, and Evolution of Magnetic Flux Ropes from the Perspective of Magnetic Twist}",
      journal = {\apj},
         year = 2016,
        month = feb,
       volume = {818},
       number = {2},
          eid = {148},
        pages = {148},
          doi = {10.3847/0004-637X/818/2/148},
archivePrefix = {arXiv},
       eprint = {1512.02338},
 primaryClass = {astro-ph.SR},
       adsurl = {https://ui.adsabs.harvard.edu/abs/2016ApJ...818..148L}
}

@ARTICLE{chitta2025,
       author = {{Chitta}, L.~P. and {Huang}, Z. and {D'Amicis}, R. and {Calchetti}, D. and {Zhukov}, A.~N. and {Kraaikamp}, E. and {Verbeeck}, C. and {Aznar Cuadrado}, R. and {Hirzberger}, J. and {Berghmans}, D. and {Horbury}, T.~S. and {Solanki}, S.~K. and {Owen}, C.~J. and {Harra}, L. and {Peter}, H. and {Sch{\"u}hle}, U. and {Teriaca}, L. and {Louarn}, P. and {Livi}, S. and {Giunta}, A.~S. and {Hassler}, D.~M. and {Wang}, Y. -M.},
        title = "{Coronal hole picoflare jets are progenitors of both fast and Alfv{\'e}nic slow solar wind}",
      journal = {\aap},
         year = 2025,
        month = feb,
       volume = {694},
          eid = {A71},
        pages = {A71},
          doi = {10.1051/0004-6361/202452737},
archivePrefix = {arXiv},
       eprint = {2411.16513},
 primaryClass = {astro-ph.SR},
       adsurl = {https://ui.adsabs.harvard.edu/abs/2025A&A...694A..71C}
}

@ARTICLE{chen2021,
       author = {{Chen}, Yajie and {Przybylski}, Damien and {Peter}, Hardi and {Tian}, Hui and {Auch{\`e}re}, F. and {Berghmans}, D.},
        title = "{Transient small-scale brightenings in the quiet solar corona: A model for campfires observed with Solar Orbiter}",
      journal = {\aap},
         year = 2021,
        month = dec,
       volume = {656},
          eid = {L7},
        pages = {L7},
          doi = {10.1051/0004-6361/202140638},
archivePrefix = {arXiv},
       eprint = {2104.10940},
 primaryClass = {astro-ph.SR},
       adsurl = {https://ui.adsabs.harvard.edu/abs/2021A&A...656L...7C}
}

@ARTICLE{dammasch1999,
       author = {{Dammasch}, I.~E. and {Wilhelm}, K. and {Curdt}, W. and {Hassler}, D.~M.},
        title = "{The NE BT VIII (lambda 770) resonance line: solar wavelengths determined by SUMER on SOHO}",
      journal = {\aap},
         year = 1999,
        month = jun,
       volume = {346},
        pages = {285-294},
       adsurl = {https://ui.adsabs.harvard.edu/abs/1999A&A...346..285D}
}

@ARTICLE{hassler1999,
       author = {{Hassler}, Donald M. and {Dammasch}, Ingolf E. and {Lemaire}, Philippe and {Brekke}, Pal and {Curdt}, Werner and {Mason}, Helen E. and {Vial}, Jean-Claude and {Wilhelm}, Klaus},
        title = "{Solar Wind Outflow and the Chromospheric Magnetic Network}",
      journal = {Science},
         year = 1999,
        month = feb,
       volume = {283},
        pages = {810},
          doi = {10.1126/science.283.5403.810},
       adsurl = {https://ui.adsabs.harvard.edu/abs/1999Sci...283..810H}
}

@ARTICLE{milanovic2023,
       author = {{Milanovi{\'c}}, N. and {Chitta}, L.~P. and {Peter}, H.},
        title = "{Diffuse solar coronal features and their spicular footpoints}",
      journal = {\aap},
         year = 2023,
        month = may,
       volume = {673},
          eid = {A81},
        pages = {A81},
          doi = {10.1051/0004-6361/202245544},
archivePrefix = {arXiv},
       eprint = {2303.13161},
 primaryClass = {astro-ph.SR},
       adsurl = {https://ui.adsabs.harvard.edu/abs/2023A&A...673A..81M}
}

@ARTICLE{suetterlin1998,
       author = {{Suetterlin}, Peter},
        title = "{Properties of solar pores}",
      journal = {\aap},
         year = 1998,
        month = may,
       volume = {333},
        pages = {305-312},
       adsurl = {https://ui.adsabs.harvard.edu/abs/1998A&A...333..305S}
}

@ARTICLE{depontieu2007,
       author = {{De Pontieu}, B. and {McIntosh}, S.~W. and {Carlsson}, M. and {Hansteen}, V.~H. and {Tarbell}, T.~D. and {Schrijver}, C.~J. and {Title}, A.~M. and {Shine}, R.~A. and {Tsuneta}, S. and {Katsukawa}, Y. and {Ichimoto}, K. and {Suematsu}, Y. and {Shimizu}, T. and {Nagata}, S.},
        title = "{Chromospheric Alfv{\'e}nic Waves Strong Enough to Power the Solar Wind}",
      journal = {Science},
         year = 2007,
        month = dec,
       volume = {318},
       number = {5856},
        pages = {1574},
          doi = {10.1126/science.1151747},
       adsurl = {https://ui.adsabs.harvard.edu/abs/2007Sci...318.1574D}
}

@ARTICLE{McIntosh2011,
       author = {{McIntosh}, Scott W. and {de Pontieu}, Bart and {Carlsson}, Mats and {Hansteen}, Viggo and {Boerner}, Paul and {Goossens}, Marcel},
        title = "{Alfv{\'e}nic waves with sufficient energy to power the quiet solar corona and fast solar wind}",
      journal = {\nat},
         year = 2011,
        month = jul,
       volume = {475},
       number = {7357},
        pages = {477-480},
          doi = {10.1038/nature10235},
       adsurl = {https://ui.adsabs.harvard.edu/abs/2011Natur.475..477M}
}

@ARTICLE{rempel2014,
       author = {{Rempel}, M.},
        title = "{Numerical Simulations of Quiet Sun Magnetism: On the Contribution from a Small-scale Dynamo}",
      journal = {\apj},
         year = 2014,
        month = jul,
       volume = {789},
       number = {2},
          eid = {132},
        pages = {132},
          doi = {10.1088/0004-637X/789/2/132},
archivePrefix = {arXiv},
       eprint = {1405.6814},
 primaryClass = {astro-ph.SR},
       adsurl = {https://ui.adsabs.harvard.edu/abs/2014ApJ...789..132R}
}

@ARTICLE{Priest2002,
       author = {{Priest}, Eric R. and {Heyvaerts}, Jean F. and {Title}, Alan M.},
        title = "{A Flux-Tube Tectonics Model for Solar Coronal Heating Driven by the Magnetic Carpet}",
      journal = {\apj},
         year = 2002,
        month = sep,
       volume = {576},
       number = {1},
        pages = {533-551},
          doi = {10.1086/341539},
       adsurl = {https://ui.adsabs.harvard.edu/abs/2002ApJ...576..533P}
}

@ARTICLE{Matthaeus1999,
       author = {{Matthaeus}, W.~H. and {Zank}, G.~P. and {Oughton}, S. and {Mullan}, D.~J. and {Dmitruk}, P.},
        title = "{Coronal Heating by Magnetohydrodynamic Turbulence Driven by Reflected Low-Frequency Waves}",
      journal = {\apjl},
         year = 1999,
        month = sep,
       volume = {523},
       number = {1},
        pages = {L93-L96},
          doi = {10.1086/312259},
       adsurl = {https://ui.adsabs.harvard.edu/abs/1999ApJ...523L..93M}
}

@ARTICLE{Leamon2000,
       author = {{Leamon}, R.~J. and {Matthaeus}, W.~H. and {Smith}, C.~W. and {Zank}, G.~P. and {Mullan}, D.~J. and {Oughton}, S.},
        title = "{MHD-driven Kinetic Dissipation in the Solar Wind and Corona}",
      journal = {\apj},
         year = 2000,
        month = jul,
       volume = {537},
       number = {2},
        pages = {1054-1062},
          doi = {10.1086/309059},
       adsurl = {https://ui.adsabs.harvard.edu/abs/2000ApJ...537.1054L}
}

@ARTICLE{Dmitruk2002,
       author = {{Dmitruk}, P. and {Matthaeus}, W.~H. and {Milano}, L.~J. and {Oughton}, S. and {Zank}, G.~P. and {Mullan}, D.~J.},
        title = "{Coronal Heating Distribution Due to Low-Frequency, Wave-driven Turbulence}",
      journal = {\apj},
         year = 2002,
        month = aug,
       volume = {575},
       number = {1},
        pages = {571-577},
          doi = {10.1086/341188},
archivePrefix = {arXiv},
       eprint = {astro-ph/0204347},
 primaryClass = {astro-ph},
       adsurl = {https://ui.adsabs.harvard.edu/abs/2002ApJ...575..571D}
}

@ARTICLE{Servidio2010,
       author = {{Servidio}, S. and {Matthaeus}, W.~H. and {Shay}, M.~A. and {Dmitruk}, P. and {Cassak}, P.~A. and {Wan}, M.},
        title = "{Statistics of magnetic reconnection in two-dimensional magnetohydrodynamic turbulence}",
      journal = {Physics of Plasmas},
         year = 2010,
        month = mar,
       volume = {17},
       number = {3},
          eid = {032315},
        pages = {032315},
          doi = {10.1063/1.3368798},
       adsurl = {https://ui.adsabs.harvard.edu/abs/2010PhPl...17c2315S}
}

@ARTICLE{Rappazzo2012,
       author = {{Rappazzo}, A.~F. and {Matthaeus}, W.~H. and {Ruffolo}, D. and {Servidio}, S. and {Velli}, M.},
        title = "{Interchange Reconnection in a Turbulent Corona}",
      journal = {\apjl},
         year = 2012,
        month = oct,
       volume = {758},
       number = {1},
          eid = {L14},
        pages = {L14},
          doi = {10.1088/2041-8205/758/1/L14},
archivePrefix = {arXiv},
       eprint = {1209.1388},
 primaryClass = {astro-ph.SR},
       adsurl = {https://ui.adsabs.harvard.edu/abs/2012ApJ...758L..14R}
}

@ARTICLE{Vidotto2021LRSP,
       author = {{Vidotto}, Aline A.},
        title = "{The evolution of the solar wind}",
      journal = {Living Reviews in Solar Physics},
         year = 2021,
        month = dec,
       volume = {18},
       number = {1},
          eid = {3},
        pages = {3},
          doi = {10.1007/s41116-021-00029-w},
archivePrefix = {arXiv},
       eprint = {2103.15748},
 primaryClass = {astro-ph.SR},
       adsurl = {https://ui.adsabs.harvard.edu/abs/2021LRSP...18....3V}
}

@ARTICLE{Prsa2016AJ,
       author = {{Pr{\v{s}}a}, Andrej and {Harmanec}, Petr and {Torres}, Guillermo and {Mamajek}, Eric and {Asplund}, Martin and {Capitaine}, Nicole and {Christensen-Dalsgaard}, J{\o}rgen and {Depagne}, {\'E}ric and {Haberreiter}, Margit and {Hekker}, Saskia and {Hilton}, James and {Kopp}, Greg and {Kostov}, Veselin and {Kurtz}, Donald W. and {Laskar}, Jacques and {Mason}, Brian D. and {Milone}, Eugene F. and {Montgomery}, Michele and {Richards}, Mercedes and {Schmutz}, Werner and {Schou}, Jesper and {Stewart}, Susan G.},
        title = "{Nominal Values for Selected Solar and Planetary Quantities: IAU 2015 Resolution B3}",
      journal = {\aj},
         year = 2016,
        month = aug,
       volume = {152},
       number = {2},
          eid = {41},
        pages = {41},
          doi = {10.3847/0004-6256/152/2/41},
archivePrefix = {arXiv},
       eprint = {1605.09788},
 primaryClass = {astro-ph.SR},
       adsurl = {https://ui.adsabs.harvard.edu/abs/2016AJ....152...41P}
}

@ARTICLE{HaynesParnell2007,
       author = {{Haynes}, A.~L. and {Parnell}, C.~E.},
        title = "{A trilinear method for finding null points in a three-dimensional vector space}",
      journal = {Physics of Plasmas},
         year = 2007,
        month = aug,
       volume = {14},
       number = {8},
        pages = {082107-082107},
          doi = {10.1063/1.2756751},
archivePrefix = {arXiv},
       eprint = {0706.0521},
 primaryClass = {astro-ph},
       adsurl = {https://ui.adsabs.harvard.edu/abs/2007PhPl...14h2107H}
}

@ARTICLE{Masson2009,
       author = {{Masson}, S. and {Pariat}, E. and {Aulanier}, G. and {Schrijver}, C.~J.},
        title = "{The Nature of Flare Ribbons in Coronal Null-Point Topology}",
      journal = {\apj},
         year = 2009,
        month = jul,
       volume = {700},
       number = {1},
        pages = {559-578},
          doi = {10.1088/0004-637X/700/1/559},
       adsurl = {https://ui.adsabs.harvard.edu/abs/2009ApJ...700..559M}
}

@ARTICLE{Pontin2016,
       author = {{Pontin}, David and {Galsgaard}, Klaus and {D{\'e}moulin}, Pascal},
        title = "{Why Are Flare Ribbons Associated with the Spines of Magnetic Null Points Generically Elongated?}",
      journal = {\solphys},
         year = 2016,
        month = aug,
       volume = {291},
       number = {6},
        pages = {1739-1759},
          doi = {10.1007/s11207-016-0919-9},
archivePrefix = {arXiv},
       eprint = {1605.05704},
 primaryClass = {astro-ph.SR},
       adsurl = {https://ui.adsabs.harvard.edu/abs/2016SoPh..291.1739P}
}

@ARTICLE{Masson2017,
       author = {{Masson}, Sophie and {Pariat}, {\'E}tienne and {Valori}, Gherardo and {Deng}, Na and {Liu}, Chang and {Wang}, Haimin and {Reid}, Hamish},
        title = "{Flux rope, hyperbolic flux tube, and late extreme ultraviolet phases in a non-eruptive circular-ribbon flare}",
      journal = {\aap},
         year = 2017,
        month = aug,
       volume = {604},
          eid = {A76},
        pages = {A76},
          doi = {10.1051/0004-6361/201629654},
archivePrefix = {arXiv},
       eprint = {1704.01450},
 primaryClass = {astro-ph.SR},
       adsurl = {https://ui.adsabs.harvard.edu/abs/2017A&A...604A..76M}
}

@ARTICLE{DeForest1997,
       author = {{DeForest}, C.~E. and {Hoeksema}, J.~T. and {Gurman}, J.~B. and {Thompson}, B.~J. and {Plunkett}, S.~P. and {Howard}, R. and {Harrison}, R.~C. and {Hasslerz}, D.~M.},
        title = "{Polar Plume Anatomy: Results of a Coordinated Observation}",
      journal = {\solphys},
         year = 1997,
        month = oct,
       volume = {175},
       number = {2},
        pages = {393-410},
          doi = {10.1023/A:1004955223306},
       adsurl = {https://ui.adsabs.harvard.edu/abs/1997SoPh..175..393D}
}

@ARTICLE{Avallone2018,
       author = {{Avallone}, Ellis A. and {Tiwari}, Sanjiv K. and {Panesar}, Navdeep K. and {Moore}, Ronald L. and {Winebarger}, Amy},
        title = "{Critical Magnetic Field Strengths for Solar Coronal Plumes in Quiet Regions and Coronal Holes?}",
      journal = {\apj},
         year = 2018,
        month = jul,
       volume = {861},
       number = {2},
          eid = {111},
        pages = {111},
          doi = {10.3847/1538-4357/aac82c},
archivePrefix = {arXiv},
       eprint = {1805.11188},
 primaryClass = {astro-ph.SR},
       adsurl = {https://ui.adsabs.harvard.edu/abs/2018ApJ...861..111A}
}

@ARTICLE{Tsuneta2008,
       author = {{Tsuneta}, S. and {Ichimoto}, K. and {Katsukawa}, Y. and {Lites}, B.~W. and {Matsuzaki}, K. and {Nagata}, S. and {Orozco Su{\'a}rez}, D. and {Shimizu}, T. and {Shimojo}, M. and {Shine}, R.~A. and {Suematsu}, Y. and {Suzuki}, T.~K. and {Tarbell}, T.~D. and {Title}, A.~M.},
        title = "{The Magnetic Landscape of the Sun's Polar Region}",
      journal = {\apj},
         year = 2008,
        month = dec,
       volume = {688},
       number = {2},
        pages = {1374-1381},
          doi = {10.1086/592226},
archivePrefix = {arXiv},
       eprint = {0807.4631},
 primaryClass = {astro-ph},
       adsurl = {https://ui.adsabs.harvard.edu/abs/2008ApJ...688.1374T}
}

@ARTICLE{Chen2025,
       author = {{Chen}, Yajie and {Peter}, Hardi and {Przybylski}, Damien and {Iijima}, Haruhisa and {Chitta}, Lakshmi Pradeep},
        title = "{Magnetic reconnection sustains the mass budget of the solar wind}",
      journal = {\aap},
         year = 2025,
        month = oct,
       volume = {702},
          eid = {L4},
        pages = {L4},
          doi = {10.1051/0004-6361/202556696},
archivePrefix = {arXiv},
       eprint = {2509.11692},
 primaryClass = {astro-ph.SR},
       adsurl = {https://ui.adsabs.harvard.edu/abs/2025A&A...702L...4C}
}
\bibliographystyle{aasjournal}

\end{document}